\documentclass[twocolumn]{revtex4}
\usepackage{amsmath}
\usepackage{amsfonts}
\usepackage{mathbbold}
\usepackage{amssymb}
\usepackage{charter}
\usepackage{graphicx}

\begin{document}

\title{Exploring entanglement, Wigner negativity and Bell nonlocality for
anisotropic two-qutrit states}
\author{Huan Liu$^{1}$, Zu-wu Chen$^{1}$, Xue-feng Zhan$^{1}$, Hong-chun Yuan%
$^{2}$ and Xue-xiang Xu$^{1,\dag }$}
\affiliation{$^{1}$College of Physics and Communication Electronics, Jiangxi Normal
University, Nanchang 330022, China;\\
$^{2}$School of Electrical and Optoelectronic Engineering, Changzhou
Institute of Technology, Changzhou 213032, China;\\
$^{\dag }$xuxuexiang@jxnu.edu.cn}

\begin{abstract}
We introduce a family of anisotropic two-qutrit states (AITTSs). These
AITTSs are expressed as $\rho _{aiso}=p\left\vert \psi _{\left( \theta ,\phi
\right) }\right\rangle \left\langle \psi _{\left( \theta ,\phi \right)
}\right\vert +(1-p)\frac{1_{9}}{9}$ with $\left\vert \psi _{\left( \theta
,\phi \right) }\right\rangle =\sin \theta \cos \phi \left\vert
00\right\rangle +\sin \theta \sin \phi \left\vert 11\right\rangle +\cos
\theta \left\vert 22\right\rangle $ and $1_{9}=\sum_{j,k=0}^{2}\left\vert
jk\right\rangle \left\langle jk\right\vert $. For a given $p\in \lbrack 0,1]$%
, these states are adjustable in different ($\theta ,\phi $) directions. In
the case of ($\theta ,\phi $) = ($\arccos (1/\sqrt{3}),\pi /4$), the AITTS
will reduce to the isotropic two-qutrit state $\rho _{iso}$. In addition,
the AITTSs are severely affected by the white noise ($\rho _{noise}=\frac{%
1_{9}}{9}$). Three properties of the AITTSs, including entanglement, Wigner
negativity and Bell nonlocality, are explored detailedly in the analytical
and numerical ways. Each property is witnessed by an appropriate existing
criterion. Some of our results are summarized as follows: (i) Large
entanglement does not necessarily mean high Wigner negativity and strong
Bell nonlocality. (ii) A pure state with a large Schmidt number does not
necessarily have a greater Wigner negativity. (iii) Only when $\left\vert
\psi _{\left( \theta ,\phi \right) }\right\rangle $ has the Schmidt number
3, the AITTS has the possibility of exhibiting Bell nonlocality in proper
parameter range.

\textbf{Keywords:}
\end{abstract}

\maketitle

\section{ Introduction}

In various quantum fields, it is necessary to utilize quantum resources to
leverage their advantages. Then, what are quantum resources?\cite{1,2} All
those quantum properties, we think, having the ability of going beyond
classical ones in performing technological tasks, can be regarded as quantum
resources. Many properties, such as nonclassicality\cite{3,4,5,6},
non-Gaussianity\cite{7,8,9,10}, entanglement\cite{11,12,13}, steering\cite%
{14,15,16}, Bell nonlocality\cite{17,18}, Wigner negativity\cite{9,19,20},
contextuality\cite{21,22,23,24,25}, and so on, are the quantum resources.
Many researchers have studied theoretical advantages and experimental
applications of theses properties. Of course, all these properties have
their respective certification/quantification ways and are somewhat
correlated with each other\cite{26,27,28,29,30}.

Mathematically, quantum states describing quantum systems can be represented
in various ways such as state vectors, density operators, wave functions,
etc \cite{31,32}. As a fundamental tool in quantum mechanics and quantum
optics, Wigner function\cite{33,34,35} provides a quantum phase-space
representation for quantum state in terms of position and momentum,
analogous to classical phase space. However, Wigner function is often called
as a quasi-probability distribution because it can take negative values.
This feature can be quantified by the volumn of the negative part, i.e.,
Wigner negativity\cite{36,37,38}. Physically, Wigner negativity is a
rigorous non-classical maker of quantum states\cite{39}, which reveals
intrinsically non-classical behaviors (e.g., superposition, interference and
entanglement)\cite{40,41,42}. Some studies have reported that Wigner
negativity is the necessary resource for quantum computing\cite{43,44}.

As Hudson's theorem\cite{45} established, for a continuous-variable system,
the Wigner function of a pure state is non-negative if and only if it is a
Gaussian state. While the Wigner function of a mixed state is non-negative
if and only if it is a convex mixture of Gaussian states\cite{46}. Hudson's
theorem was extended into finite-dimensional systems by Gross\cite{47}. They
showed that, the Wigner function of a pure state is non-negative if and only
if it is a stabilizer state. While the Wigner function of a mixed state is
non-negative if and only if it is a convex mixture of stabilizer states\cite%
{48}. In these ways, we can understand that what kind of quantum states can
exhibit Wigner negativities.

One can distinguish continuous-variable from discrete-variable quantum state
by observing the Wigner function. Compared to the discrete case, people are
more familiar with continuous Wigner function. With the development of
quantum information, people have become increasingly enthusiastic about
studying discrete Wigner functions (DWFs) in the past two decades. The DWFs
have become useful tools of studying finite-dimensional quantum states\cite%
{49,50}. In 2004, Gibbon, Hoffman, and Wootter developed the Wigner
functions and investigated a class of DWFs\cite{51}. Subsequently, Galvao
conjectured that the discrete Wigner negativity was necessary for quantum
computation speedup\cite{52}. In 2017, Kocia and Love studied the DWFs for
qubits\cite{53}. In 2024, Wootters studied the DWFs for two-qubit states, in
order to interpret symplectic linear transformation in phase space\cite{54}.
Recently, Antonopoulos and his co-workers presented a grand unification for
all DWFs\cite{55}.

It is well known that, the qubit, as a two-state (or two-level) system, is
the basic storage unit of quantum information\cite{56}. However, more and
more practical quantum protocols require high-dimensional storage units\cite%
{57,58}. This trend triggers considerable researches related with the
qudits. As the name suggests, the qudit is the $d$-state (or $d$-level)
physical system, corresponding to $d$-dimensional mathematical model. For
some technological tasks, qudits perhaps may be more efficient than qubits.
In the current era, the internet has become indispensable in our daily
lives. Subsequently, the quantum internet\cite{59,60,61} came into being,
which has aroused extensive research interests of scientists. The main
characteristic of quantum internet is to distribute and share information
among many sites at a certain distance. Therefore, quantum internet must be
realized in multipartite scenarios. Above mentioned reasons are driving the
advances in multipartite and high-dimensional systems\cite{62,63,64}.

Every knows that entanglement is the crucial resource to achieve quantum
advantageous. In recent years, many groups are dedicated to studying the
entanglement for high-dimensional systems\cite{65,66,67,68,69,70}. In
addition, Bell nonlocality becomes another current hot research topic. Since
Bell proposed the original idea of using inequalities to witness nonlocality%
\cite{71}, many researchers have conducted extensive researches on
nonlocality. Most of works focus on two aspects: one is to construct
different inequalities by changing measurement scenarios\cite{72,73,74}, and
the other is to explore the Bell nonlocality for various multipartite and
high-dimensional quantum systems\cite{75,76}. Recently, Fonseca and his
co-workers made a survey the Bell nonlocality of entangled qudits\cite{77}.
In this regard, we particularly emphasize that, Collins, Gisin, Linden,
Massar, and Popescu developed an approach to construct Bell inequalities for
any bipartite high-dimensional quantum systems\cite{78}. These
approach-related Bell inequalities were called the CGLMP-inequalities by
later researchers. In the context of the CGLMP-inequalities, many
researchers have conducted a large number of studies on Bell non-locality%
\cite{79,80,81,82}.

As the simplest model of the multipartite and high-dimensional systems,
two-qutrit states are often chosen as examples to conduct researches on
quantum properties\cite{83,84}. In fact, two-qutrit states are just
bipartite three-dimensional states, which can be used in various physical
platforms\cite{85,86,87}. In 2012, Gruca, Laskowski, and Zukowski reported
the nonclassicality for pure two-qutrit entangled states\cite{88}. On the
other hand, noises inevitably affects the properties of quantum states. For
instance, Roy and his co-workers found that the white noise will affect the
robustness of higher-dimensional nonlocality\cite{89}. Lifshitz compared\
and analyzed noise-robustness in various self-testing protocols\cite{90}.

Combinating pure two-qutrit states with white noises, we introduce a family
of anisotropic two-qutrit states (AITTSs), which are the extension of the
isotropic two-qutrit state. To the best of our knowledge, these AITTSs and
their detailed properties are not studied completely in previous works. We
will explore entanglement, Wigner negativity and Bell-nonlcality for the
AITTSs. The paper is organized as follows: In Sec.II, we introduce the
AITTSs. In Sec.III, we analyze their entanglement in terms of an appropriate
witness. In Sec.IV, we analyze their DWFs, and then study their Wigner
negativities. In Sec.V, we study their Bell nonlocality, by checking the
violation of the CGLMP inequality. We conclude in the last section.

\section{Anisotropic two-qutrit states}

A single-qutrit state can be described in the Hilbert space spanned by three
bases $\{\left\vert 0\right\rangle ,\left\vert 1\right\rangle ,\left\vert
2\right\rangle \}$, with $\left\vert 0\right\rangle =(%
\begin{array}{ccc}
1 & 0 & 0%
\end{array}%
)^{T}$, $\left\vert 1\right\rangle =(%
\begin{array}{ccc}
0 & 1 & 0%
\end{array}%
)^{T}$, and $\left\vert 2\right\rangle =(%
\begin{array}{ccc}
0 & 0 & 1%
\end{array}%
)^{T}$. Consequently, a two-qutrit state can be described in the
nine-dimensional space spanned by nine bases, i.e., \{$\left\vert
00\right\rangle $, $\left\vert 01\right\rangle $, $\left\vert
02\right\rangle $, $\left\vert 10\right\rangle $, $\left\vert
11\right\rangle $, $\left\vert 12\right\rangle $, $\left\vert
20\right\rangle $, $\left\vert 21\right\rangle $, $\left\vert
22\right\rangle $\}. We assume that the two-qutrit state is shared by qutrit
A and qutrit B, with $\left\vert jk\right\rangle =\left\vert j\right\rangle
_{A}\otimes \left\vert k\right\rangle _{B}$ ($j,k\in
\mathbb{Z}
_{3}=\{0,1,2\}$). In general, pure two-qutrit states can be expressed as $%
\left\vert \psi _{pure}\right\rangle =\sum_{j,k=0}^{2}c_{jk}\left\vert
jk\right\rangle $ with $c_{jk}\in
\mathbb{C}
$ and $\sum_{j,k=0}^{2}\left\vert c_{jk}\right\vert ^{2}=1$. For instance,
Liang et al. discussed the properties for some pure two-qutrit states, such
as $(\left\vert 00\right\rangle +i\left\vert 22\right\rangle )/\sqrt{2}$, $%
(\left\vert 11\right\rangle +i\left\vert 22\right\rangle )/\sqrt{2}$, and $%
(i\left\vert 02\right\rangle +i\left\vert 12\right\rangle +\left\vert
10\right\rangle +\left\vert 12\right\rangle )/2$\cite{91}.

Many researchers have been conducted on the properties of various isotropic
two-qudit states\cite{64,65,92,93}. In the case of $d=3$, we can express the
isotropic two-qutrit state as
\begin{equation}
\rho _{iso}=p\left\vert \Phi _{3}^{+}\right\rangle \left\langle \Phi
_{3}^{+}\right\vert +(1-p)\frac{1_{9}}{9}.  \label{1.1}
\end{equation}%
Here, $\left\vert \Phi _{3}^{+}\right\rangle =\left( \left\vert
00\right\rangle +\left\vert 11\right\rangle +\left\vert 22\right\rangle
\right) /\sqrt{3}$ is the maximally entangled two-qutrit state (i.e., qutrit
Bell state). And, $\frac{1_{9}}{9}=\rho _{noise}$ denotes the two-qutrit
white noise, with the $9\times 9$\ identity matrix $1_{9}=\sum_{j,k=0}^{2}%
\left\vert jk\right\rangle \left\langle jk\right\vert $. From the form, $%
\rho _{iso}$ is a mixed state composed of $\left\vert \Phi
_{3}^{+}\right\rangle \left\langle \Phi _{3}^{+}\right\vert $\ with ratio $p$%
\ and $\rho _{noise}$\ with ratio $1-p$. In a sense, the parameter $p$
denotes is the probability that $\left\vert \Phi _{3}^{+}\right\rangle $\ is
unaffected by noise.

If $\left\vert \Phi _{3}^{+}\right\rangle $ of $\rho _{iso}$ in Eq.(\ref{1.1}%
)\ is replaced by $\left\vert \psi _{\left( \theta ,\phi \right)
}\right\rangle =\sin \theta \cos \phi \left\vert 00\right\rangle +\sin
\theta \sin \phi \left\vert 11\right\rangle +\cos \theta \left\vert
22\right\rangle $, we introduce \textbf{a}n\textbf{i}sotropic \textbf{t}wo-qu%
\textbf{t}rit \textbf{s}tates with the form%
\begin{equation}
\rho _{aiso}=p\left\vert \psi _{\left( \theta ,\phi \right) }\right\rangle
\left\langle \psi _{\left( \theta ,\phi \right) }\right\vert +(1-p)\frac{%
1_{9}}{9}.  \label{1.2}
\end{equation}%
For the convenience of writing, these states are abbreviated as AITTSs. And,
we assume that they are adjustable within $\theta \in \lbrack 0,\pi ]$\ , $%
\phi \in \lbrack 0,2\pi ]$ and $p\in \lbrack 0,1]$. Two extreme scenarios\
will happen, that is, $\rho _{aiso}\rightarrow \rho _{noise}$ if $p=0$ and $%
\rho _{aiso}\rightarrow \left\vert \psi _{\left( \theta ,\phi \right)
}\right\rangle $ if $p=1$. If $\left\vert \psi _{\left( \theta ,\phi \right)
}\right\rangle $ in Eq.(\ref{1.2})\ is further replaced by arbitrary $%
\left\vert \psi _{pure}\right\rangle $, the anisotropic character will be
stronger.

In the Hilbert space of two-qutrit systems, $\rho _{aiso}$ can be expanded as%
\begin{equation}
\rho _{aiso}=\left(
\begin{array}{ccccccccc}
\kappa _{1} & 0 & 0 & 0 & \tau _{1} & 0 & 0 & 0 & \tau _{2} \\
0 & \epsilon & 0 & 0 & 0 & 0 & 0 & 0 & 0 \\
0 & 0 & \epsilon & 0 & 0 & 0 & 0 & 0 & 0 \\
0 & 0 & 0 & \epsilon & 0 & 0 & 0 & 0 & 0 \\
\tau _{1} & 0 & 0 & 0 & \kappa _{2} & 0 & 0 & 0 & \tau _{3} \\
0 & 0 & 0 & 0 & 0 & \epsilon & 0 & 0 & 0 \\
0 & 0 & 0 & 0 & 0 & 0 & \epsilon & 0 & 0 \\
0 & 0 & 0 & 0 & 0 & 0 & 0 & \epsilon & 0 \\
\tau _{2} & 0 & 0 & 0 & \tau _{3} & 0 & 0 & 0 & \kappa _{3}%
\end{array}%
\right)  \label{1.3}
\end{equation}%
with $\epsilon =(1-p)/9$, $\kappa _{1}=p\sin ^{2}\theta \cos ^{2}\phi
+\epsilon $, $\kappa _{2}=$ $p\sin ^{2}\theta \sin ^{2}\phi +\epsilon $, $%
\kappa _{3}=$ $p\cos ^{2}\theta +\epsilon $, $\tau _{1}=(p\sin ^{2}\theta
\sin 2\phi )/2$, $\tau _{2}=(p\sin 2\theta \cos \phi )/2$, $\tau _{3}=(p\sin
2\theta \sin \phi )/2$.

For $\left\vert \psi _{\left( \theta ,\phi \right) }\right\rangle $, we
would like to give more detailed explanations. Formally, $\left\vert \psi
_{\left( \theta ,\phi \right) }\right\rangle $\ is the special case of $%
\left\vert \psi _{pure}\right\rangle $\ with $c_{00}=\sin \theta \cos \phi $%
, $c_{11}=\sin \theta \sin \phi $, $c_{22}=\cos \theta $, and $%
c_{01}=c_{02}=c_{10}=c_{12}=c_{20}=c_{21}=0$. Here, we define three
coefficients ($c_{00}$, $c_{11}$, and $c_{22}$) of $\left\vert \psi _{\left(
\theta ,\phi \right) }\right\rangle $ by referring the conversion between
spherical coordinates (Radius $r=1$, polar angle $\theta $, and azimuthal
angle $\phi $) and Cartesian coordinates ($x=c_{00}$, $y=c_{11}$, and $%
z=c_{22}$). In terms of Schmidt number(Sn)\cite{94,95,96,97} determined by
the coefficients, $\left\vert \psi _{\left( \theta ,\phi \right)
}\right\rangle $\ may be classified into the following possible Schmidt
decompositions.

(Sn-1) If there is only one non-zero coefficient, $\left\vert \psi _{\left(
\theta ,\phi \right) }\right\rangle $ will be the Sn=1 states, including $%
\left\vert \psi _{(\pi /2,0)}\right\rangle =\left\vert 00\right\rangle
\equiv \left\vert S_{1}^{(1)}\right\rangle $, $\left\vert \psi _{(\pi /2,\pi
/2)}\right\rangle =\left\vert 11\right\rangle \equiv \left\vert
S_{1}^{(2)}\right\rangle $, $\left\vert \psi _{\left( 0,\phi \right)
}\right\rangle =\left\vert 22\right\rangle \equiv \left\vert
S_{1}^{(3)}\right\rangle $.

(Sn-2) If there are two non-zero coefficients, $\left\vert \psi _{\left(
\theta ,\phi \right) }\right\rangle $ will be the Sn=2 states, including $%
\left\vert \psi _{\left( \pi /2,\phi \right) }\right\rangle =\cos \phi
\left\vert 00\right\rangle +\sin \phi \left\vert 11\right\rangle $, $%
\left\vert \psi _{\left( \theta ,0\right) }\right\rangle =\sin \theta
\left\vert 00\right\rangle +\cos \theta \left\vert 22\right\rangle $, and $%
\left\vert \psi _{\left( \theta ,\pi /2\right) }\right\rangle =\sin \theta
\left\vert 11\right\rangle +\cos \theta \left\vert 22\right\rangle $. Note
that we must ensure the condition of two non-zero coefficients. Among these
Sn=2 states, $\left\vert \psi _{(\pi /2,\pi /4)}\right\rangle =(\left\vert
00\right\rangle +\left\vert 11\right\rangle )/\sqrt{2}\equiv \left\vert
S_{2}^{(1)}\right\rangle $, $\left\vert \psi _{(\pi /4,0)}\right\rangle
=(\left\vert 00\right\rangle +\left\vert 22\right\rangle )/\sqrt{2}\equiv
\left\vert S_{2}^{(2)}\right\rangle $, and $\left\vert \psi _{(\pi /4,\pi
/2)}\right\rangle =(\left\vert 11\right\rangle +\left\vert 22\right\rangle )/%
\sqrt{2}\equiv \left\vert S_{2}^{(3)}\right\rangle $ are the maximally
entangled Sn=2 states. Others are the non-maximally entangled Sn=2 states,
such as $\left\vert \psi _{(\pi /2,\pi /6)}\right\rangle =\frac{\sqrt{3}}{2}%
\left\vert 00\right\rangle +\frac{1}{2}\left\vert 11\right\rangle )\equiv
\left\vert S_{2}^{(4)}\right\rangle $, $\left\vert \psi _{(\pi
/6,0)}\right\rangle =\frac{1}{2}\left\vert 00\right\rangle +\frac{\sqrt{3}}{2%
}\left\vert 22\right\rangle )\equiv \left\vert S_{2}^{(5)}\right\rangle $,
and $\left\vert \psi _{(\pi /6,\pi /2)}\right\rangle =\frac{1}{2}\left\vert
11\right\rangle +\frac{\sqrt{3}}{2}\left\vert 22\right\rangle \equiv
\left\vert S_{2}^{(6)}\right\rangle $.

(Sn-3) If there are three non-zero coefficients, $\left\vert \psi _{\left(
\theta ,\phi \right) }\right\rangle $ will be the Sn=3 states, such as $%
\left\vert \psi _{(\arccos (1/\sqrt{3}),\pi /4)}\right\rangle =\left\vert
\Phi _{3}^{+}\right\rangle \equiv \left\vert S_{3}^{(1)}\right\rangle $ and $%
\left\vert \psi _{(\arccos (1/\sqrt{3}),\pi /6)}\right\rangle =\frac{1}{%
\sqrt{2}}\left\vert 00\right\rangle +\frac{1}{\sqrt{6}}\left\vert
11\right\rangle +\frac{1}{\sqrt{3}}\left\vert 22\right\rangle \equiv
\left\vert S_{3}^{(2)}\right\rangle $. Since $\left\vert \psi _{\left(
\theta ,\phi \right) }\right\rangle $ will reduce to $\left\vert \Phi
_{3}^{+}\right\rangle $ if $\left( \theta ,\phi \right) =(\arccos (1/\sqrt{3}%
),\phi =\pi /4)$, $\rho _{aiso}$ in this case will reduce to $\rho _{iso}$,
together with $\kappa _{1}=\kappa _{2}=\kappa _{3}=(2p+1)/9$ and $\tau
_{1}=\tau _{2}=\tau _{3}=p/3$. It should be noted that $\left\vert
S_{3}^{(1)}\right\rangle $\ is just $\left\vert \Phi _{3}^{+}\right\rangle $%
, together with $\arccos (1/\sqrt{3})\simeq 0.955317$\ and $\pi /4\simeq
0.785398$.

In our following work, we often use above mentioned eleven states
(abbreviated the Sn=n state as $\left\vert S_{n}^{(i)}\right\rangle $) as
examples of $\left\vert \psi _{\left( \theta ,\phi \right) }\right\rangle $
to study our considered properties.

\section{Entanglement of AITTSs}

In this section, we shall quantify entanglement for AITTSs by virtue of
negativity under partial transposition\cite{98,99}. Performing partial
transposition in part A (or part B) for $\rho _{aiso}$, we obtain%
\begin{equation}
\rho _{aiso}^{T_{A}}=\rho _{aiso}^{T_{B}}=\left(
\begin{array}{ccccccccc}
\kappa _{1} & 0 & 0 & 0 & 0 & 0 & 0 & 0 & 0 \\
0 & \epsilon & 0 & \tau _{1} & 0 & 0 & 0 & 0 & 0 \\
0 & 0 & \epsilon & 0 & 0 & 0 & \tau _{2} & 0 & 0 \\
0 & \tau _{1} & 0 & \epsilon & 0 & 0 & 0 & 0 & 0 \\
0 & 0 & 0 & 0 & \kappa _{2} & 0 & 0 & 0 & 0 \\
0 & 0 & 0 & 0 & 0 & \epsilon & 0 & \tau _{3} & 0 \\
0 & 0 & \tau _{2} & 0 & 0 & 0 & \epsilon & 0 & 0 \\
0 & 0 & 0 & 0 & 0 & \tau _{3} & 0 & \epsilon & 0 \\
0 & 0 & 0 & 0 & 0 & 0 & 0 & 0 & \kappa _{3}%
\end{array}%
\right) .  \label{2.1}
\end{equation}%
The matrix of Eq.(\ref{4.1}) has nine eigenvalues $\lambda _{1}=\kappa _{1}$%
, $\lambda _{2}=\kappa _{2}$, $\lambda _{3}=\kappa _{3}$, $\lambda
_{4}=\epsilon -\tau _{1}$, $\lambda _{5}=\epsilon +\tau _{1}$, $\lambda
_{6}=\epsilon -\tau _{2}$, $\lambda _{7}=\epsilon +\tau _{2}$, $\lambda
_{8}=\epsilon -\tau _{3}$, and $\lambda _{9}=\epsilon +\tau _{3}$. Adding up
all these eigenvalues ($\sum_{j=1}^{9}\lambda _{j}$) gives $\mathrm{Tr}(\rho
_{aiso}^{T_{A}})=\mathrm{Tr}(\rho _{aiso}^{T_{B}})=1$ as expected.

The entanglement of $\rho _{aiso}$ is calculated as
\begin{equation}
\mathcal{E}\left( \rho _{aiso}\right) =\frac{1}{2}(\sum_{j=1}^{9}\left\vert
\lambda _{j}\right\vert -1),  \label{2.2}
\end{equation}%
i.e., minus sum of all negative eigenvalues ($-\sum_{i}\lambda _{i}^{-}$, $%
\lambda _{i}^{-}<0$). As references, we list the following special values
including $\mathcal{E}(\left\vert S_{1}^{(1)}\right\rangle )=\mathcal{E}%
(\left\vert S_{1}^{(2)}\right\rangle )=\mathcal{E}(\left\vert
S_{1}^{(3)}\right\rangle )=0$, $\mathcal{E}(\left\vert
S_{2}^{(1)}\right\rangle )=\mathcal{E}(\left\vert S_{2}^{(2)}\right\rangle )=%
\mathcal{E}(\left\vert S_{2}^{(3)}\right\rangle )=0.5$, $\mathcal{E}%
(\left\vert S_{2}^{(4)}\right\rangle )=\mathcal{E}(\left\vert
S_{2}^{(5)}\right\rangle )=\mathcal{E}(\left\vert S_{2}^{(6)}\right\rangle
)\simeq 0.481481$, $\mathcal{E}(\left\vert S_{3}^{(1)}\right\rangle )=1$, $%
\mathcal{E}(\left\vert S_{3}^{(2)}\right\rangle )\simeq 0.932626$, and $%
\mathcal{E}\left( \rho _{noise}\right) =0$.

Figure 1 depicts the variation of entanglement $\mathcal{E}\left( \rho
_{aiso}\right) $ versus $p$\ for eleven $(\theta ,\phi )$ cases. There are
five curves in this figure. Each curve is illustrated as follows:

(eL1) The first curve corresponds to the cases of $\left( \theta ,\phi
\right) =(\pi /2,0)$, $(\pi /2,\pi /2)$, $\left( 0,\phi \right) $. It
satisfy $\mathcal{E}\left( \rho _{aiso}\right) \equiv 0$ for any $p\in
\lbrack 0,1]$.

(eL2) The second curve corresponds to the cases of $\left( \theta ,\phi
\right) =(\pi /2,\pi /6)$, $(\pi /6,0)$, $(\pi /6,\pi /2)$. It is a
piecewise function line, satisfying $\mathcal{E}\left( \rho _{aiso}\right)
=0 $ in the interval of $0\leq p\lesssim 0.204202$ and $\mathcal{E}\left(
\rho _{aiso}\right) \simeq 0.544124p-1/9$ in the interval of $%
0.204202\lesssim p\leq 1$.

(eL3) The third curve corresponds to the cases of $\left( \theta ,\phi
\right) =(\pi /2,\pi /4)$, $(\pi /4,0)$, $(\pi /4,\pi /2)$. It is a
piecewise function line, satisfying $\mathcal{E}\left( \rho _{aiso}\right)
=0 $ in the interval of $0\leq p\leq 2/11$ and $\mathcal{E}\left( \rho
_{aiso}\right) =$ $11p/18-1/9$ in the interval of $2/11\lesssim p\leq 1$.

(eL4) The fourth curve corresponds to the case of $\left( \theta ,\phi
\right) =(\arccos (1/\sqrt{3}),\pi /6)$. It is also a piecewise function
line, satisfying $\mathcal{E}\left( \rho _{aiso}\right) =0$ in the interval
of $0\leq p\lesssim 0.213939$, $\mathcal{E}\left( \rho _{aiso}\right) \simeq
$ $0.519359p-1/9$ in the interval of $0.213939\lesssim p\lesssim 0.277926$, $%
\mathcal{E}\left( \rho _{aiso}\right) \simeq 0.919146p-2/9$ in the interval
of $0.277926\lesssim p\lesssim 0.320377$, and $\mathcal{E}\left( \rho
_{aiso}\right) \simeq $ $1.26596p-1/3$ in the interval of $0.320377\lesssim
p\leq 1$.

(eL5) The fifth curve corresponds to the case of $\left( \theta ,\phi
\right) =(\arccos (1/\sqrt{3}),\pi /4)$. It is also a piecewise function
line, satisfying $\mathcal{E}\left( \rho _{aiso}\right) =0$ in the interval
of $0\leq p\lesssim 0.25$ and $\mathcal{E}\left( \rho _{aiso}\right) \simeq
4p/3-1/3$ in the interval of $0.25\lesssim p\leq 1$.

From above numerical results, we can infer that, for different $\left(
\theta ,\phi \right) $ cases, there are different evolution curves of $%
\mathcal{E}\left( \rho _{aiso}\right) $ over $p$. When $\left\vert \psi
_{\left( \theta ,\phi \right) }\right\rangle $ is the Sn=1 state, $\mathcal{E%
}\left( \rho _{aiso}\right) $ remains at zero in the whole $p$ range. When $%
\left\vert \psi _{\left( \theta ,\phi \right) }\right\rangle $ is not the
Sn=1 state, $\mathcal{E}\left( \rho _{aiso}\right) $ will change with the $p$%
-value. For each $\left( \theta ,\phi \right) $ case, there will be a $p$%
-value range satisfying $\mathcal{E}\left( \rho _{aiso}\right) =0$. That is
to say, only when $p$-value exceeds a certain threshold, it is possible to
observe $\mathcal{N}\left( \rho _{aiso}\right) >0$ for a given $\left(
\theta ,\phi \right) $ case. This can be seen from Fig.2, which depicts the
feasibility regions satisfying $\mathcal{E}\left( \rho _{aiso}\right) >0$ in
$\left( \theta ,\phi ,p\right) $\ space.

\begin{figure}[tbp]
\label{FigEN1} \centering\includegraphics[width=0.9\columnwidth]{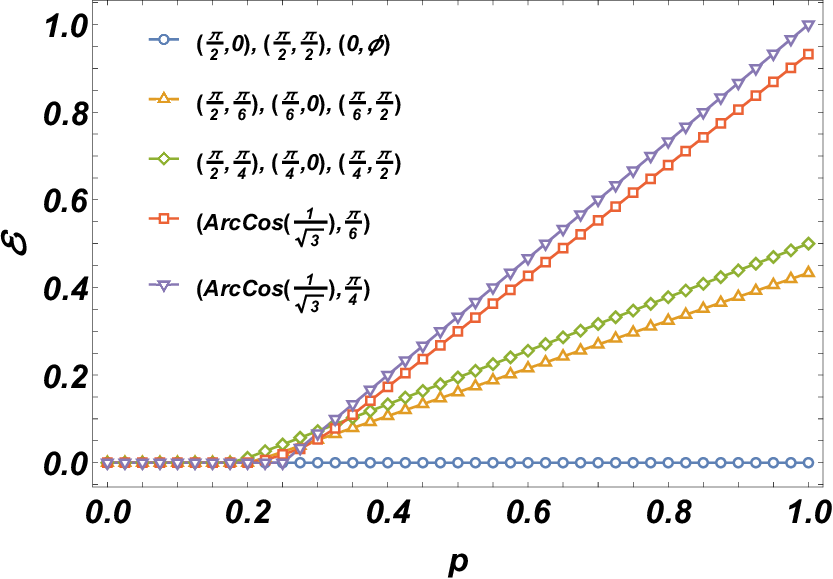}
\caption{$\mathcal{E}\left( \protect\rho _{aiso}\right) $ versus $p$\ for
eleven ($\protect\theta $, $\protect\phi $) cases. There are only five
variation curves. For $p=1$, $\mathcal{E}\left( \protect\rho _{aiso}\right) $
values are$\ 0$, $0.433013$, $0.5$, $0.932626$, $1$ in sequence.}
\end{figure}

\begin{figure}[tbp]
\label{FigEN2} \centering\includegraphics[width=0.8\columnwidth]{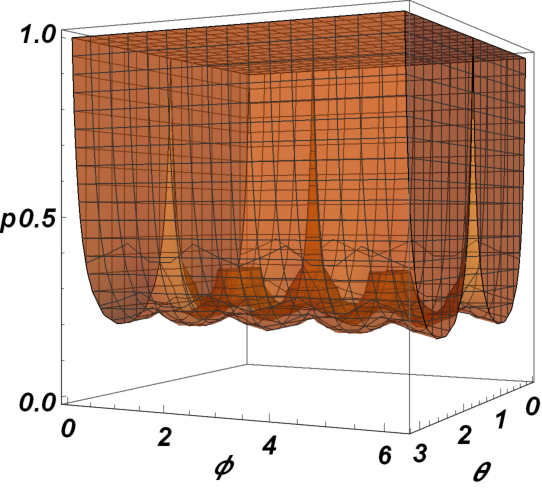}
\caption{Three-dimensional feasibility region of $\mathcal{E}\left( \protect%
\rho _{aiso}\right) >0$ showing entanglement in $(\protect\theta ,\protect%
\phi ,p)$\ space. The blank region is that satisfying $\mathcal{E}\left(
\protect\rho _{aiso}\right) =0$.}
\end{figure}

As expected, we further verify that the maximum entanglement value ($%
\mathcal{E}^{\max }\left( \rho _{aiso}\right) =1$) is positioned at $\left(
\theta ,\phi ,p\right) \simeq \left( 0.955317,0.785398,1\right) $, which
corresponds exactly to the maximum entangled state $\left\vert \Phi
_{3}^{+}\right\rangle $, i.e. $\mathcal{E}(\left\vert \Phi
_{3}^{+}\right\rangle )=1$.

\section{Wigner negativity of AITTSs}

In this section, we shall analyze the DWFs and study Wigner negativities for
the AITTSs. Regarding the foundations of this section, one can refer to two
relevant works from Delfose's group\cite{100} and Meyer's group\cite{101}.

\subsection{Discrete Wigner function}

Similar to the qubit Pauli operators $\sigma _{x}=\left(
\begin{array}{cc}
0 & 1 \\
1 & 0%
\end{array}%
\right) $\ and $\sigma _{z}=\left(
\begin{array}{cc}
1 & 0 \\
0 & -1%
\end{array}%
\right) $, one can introduce the qutrit Pauli operators%
\begin{equation}
X=\sum_{k=0}^{2}\left\vert k+1\right\rangle \left\langle k\right\vert
=\left(
\begin{array}{ccc}
0 & 0 & 1 \\
1 & 0 & 0 \\
0 & 1 & 0%
\end{array}%
\right) ,  \label{3.1}
\end{equation}%
and%
\begin{equation}
Z=\sum_{k=0}^{2}\omega ^{k}\left\vert k\right\rangle \left\langle
k\right\vert =\left(
\begin{array}{ccc}
1 & 0 & 0 \\
0 & \omega & 0 \\
0 & 0 & \omega ^{2}%
\end{array}%
\right) ,  \label{3.2}
\end{equation}%
with $\omega =e^{\frac{2\pi i}{3}}$. They obey $X\left\vert k\right\rangle
=\left\vert \left( k+1\right) \text{ mod }3\right\rangle $, $Z\left\vert
k\right\rangle =\omega ^{k}\left\vert k\right\rangle $, and $%
Z^{z}X^{x}=\omega ^{xz}X^{x}Z^{z}$ for $x,z,k\in
\mathbb{Z}
_{3}$.

For a qutrit, the Wiger operator in phase point $\left( u_{x},u_{z}\right) $
is defined as%
\begin{equation}
A_{\left( u_{x},u_{z}\right) }=\frac{1}{3}\sum_{v_{x}=0}^{2}%
\sum_{v_{z}=0}^{2}\omega ^{u_{z}v_{x}-u_{x}v_{z}}D_{\left(
v_{x},v_{z}\right) },  \label{3.3}
\end{equation}%
with the Heisenberg-Weyl displacement operator $D_{\left( x,z\right)
}=\omega ^{\frac{1}{2}xz}X^{x}Z^{z}$. Therefore, we rewrite $A_{\left(
u_{x},u_{z}\right) }$ as $A_{\left( x,z\right) }$ with the following matrix%
\begin{widetext}
\begin{equation}
A_{(x,z)}=\frac{1}{3}\allowbreak \left(
\begin{array}{ccc}
1+\omega ^{-x}+\omega ^{-2x} & \omega ^{2z-x+2}+\omega ^{2z-2x+4}+\omega
^{2z} & \omega ^{z}+\omega ^{z-2x+2}+\omega ^{z-x+\frac{5}{2}} \\
\omega ^{z}+\omega ^{z-2x+1}+\omega ^{z-x+\frac{1}{2}} & 1+\omega
^{2-2x}+\omega ^{-x+1} & \omega ^{2z-x+3}+\omega ^{2z-2x+3}+\omega ^{2z} \\
\omega ^{2z-x+1}+\omega ^{2z-2x+2}+\omega ^{2z} & \omega ^{z}+\omega
^{z-2x+3}+\omega ^{z-x+\frac{3}{2}} & 1+\omega ^{1-2x}+\omega ^{-x+2}%
\end{array}%
\right) .  \label{3.4}
\end{equation}%
\end{widetext}

For the AITTSs with the density matrix $\rho _{aiso}$, the corresponding DWF
can be calculated by%
\begin{equation}
W_{\left( x_{1},z_{1};x_{2},z_{2}\right) }\left( \rho _{aiso}\right) =\frac{1%
}{3^{2}}\mathrm{Tr}[(A_{\left( x_{1},z_{1}\right) }\otimes A_{\left(
x_{2},z_{2}\right) })\rho _{aiso}].  \label{3.5}
\end{equation}%
For every two-qutrit state, there are eighty-one phase points due to $\left(
x_{1},z_{1};x_{2},z_{2}\right) \in
\mathbb{Z}
_{3}^{4}$. In this paper, we will arrange them as follows
\begin{widetext}
\begin{equation}
\begin{array}{ccccccccc}
W_{0000} & W_{0100} & W_{0200} & W_{0001} & W_{0101} & W_{0201} & W_{0002} &
W_{0102} & W_{0202} \\
W_{1000} & W_{1100} & W_{1200} & W_{1001} & W_{1101} & W_{1201} & W_{1002} &
W_{1102} & W_{1202} \\
W_{2000} & W_{2100} & W_{2200} & W_{2001} & W_{2101} & W_{2201} & W_{2002} &
W_{2102} & W_{2202} \\
W_{0010} & W_{0110} & W_{0210} & W_{0011} & W_{0111} & W_{0211} & W_{0012} &
W_{0112} & W_{0212} \\
W_{1010} & W_{1110} & W_{1210} & W_{1011} & W_{1111} & W_{1211} & W_{1012} &
W_{1112} & W_{1212} \\
W_{2010} & W_{2110} & W_{2210} & W_{2011} & W_{2111} & W_{2211} & W_{2012} &
W_{2112} & W_{2212} \\
W_{0020} & W_{0120} & W_{0220} & W_{0021} & W_{0121} & W_{0221} & W_{0022} &
W_{0122} & W_{0222} \\
W_{1020} & W_{1120} & W_{1220} & W_{1021} & W_{1121} & W_{1221} & W_{1022} &
W_{1122} & W_{1222} \\
W_{2020} & W_{2120} & W_{2220} & W_{2021} & W_{2121} & W_{2221} & W_{2022} &
W_{2122} & W_{2222}%
\end{array}%
.  \label{3.6}
\end{equation}%
\end{widetext}Note that $W_{x_{1}z_{1}x_{2}z_{2}}\equiv W_{\left(
x_{1},z_{1};x_{2},z_{2}\right) }$. According to Eqs.(\ref{3.5}) and (\ref%
{3.6}), we can obtain the DWF values and plot the DWF figures for $\rho
_{aiso}$.

In Fig.3, we plot the DWFs for $\left\vert S_{1}^{(1)}\right\rangle $, $%
\left\vert S_{1}^{(2)}\right\rangle $, and $\left\vert
S_{1}^{(3)}\right\rangle $. Their DWFs all have nine points with value $1/9$%
\ and seventy-two points with value $0$, i.e. \{$\frac{1}{9}\rightarrow 9$, $%
0\rightarrow 72$\}. Moreover, all of their values are non-negative.

In Fig.4, we plot the DWFs for $\left\vert S_{2}^{(1)}\right\rangle $, $%
\left\vert S_{2}^{(2)}\right\rangle $, and $\left\vert
S_{2}^{(3)}\right\rangle $. Their DWFs all have three points with value $%
5/81 $, six points with value $-5/162$, twenty-four points with value $1/162$%
, twenty-four points with value $-1/81$, twelve points with value $7/162$,
six points with value $13/162$,\ and six points with value $2/81$, i.e. \{$%
\frac{5}{81}\rightarrow 3$, $-\frac{5}{162}\rightarrow 6$, $\frac{1}{162}%
\rightarrow 24$, $-\frac{1}{81}\rightarrow 24$, $\frac{7}{162}\rightarrow 12$%
, $\frac{13}{162}\rightarrow 6$, $\frac{2}{81}\rightarrow 6$\}. For these
three states, their DWFs may be negative.

In Fig.5, we plot the DWFs for $\left\vert S_{3}^{(1)}\right\rangle $ (see
left sub-figure) and $\rho _{noise}=1_{9}/9$ (see right sub-figure). For $%
\left\vert S_{3}^{(1)}\right\rangle $, its DWF has eight points with value $%
5/81$, thirty-six points with value $2/81$, thirty-six points with value $%
-1/81$, and one point with value $7/162$, i.e. \{$\frac{5}{81}\rightarrow 8$%
, $\frac{2}{81}\rightarrow 36$, $-\frac{1}{81}\rightarrow 36$, $\frac{7}{162}%
\rightarrow 1$\}. That is, the DWF of $\left\vert \Phi _{3}^{+}\right\rangle
$ may be negative. For $\rho _{noise}$, its DWF has thirty-six points with
value $0$, thirty-six points with value $1/54$, and nine points with value $%
1/27$, i.e. \{$0\rightarrow 36$, $\frac{1}{54}\rightarrow 36$, $\frac{1}{27}%
\rightarrow 9$\}. That is, the DWF of $\rho _{noise}$ is non-negative.

\begin{figure*}[tbp]
\label{FigWF1} \centering\includegraphics[width=1.9\columnwidth]{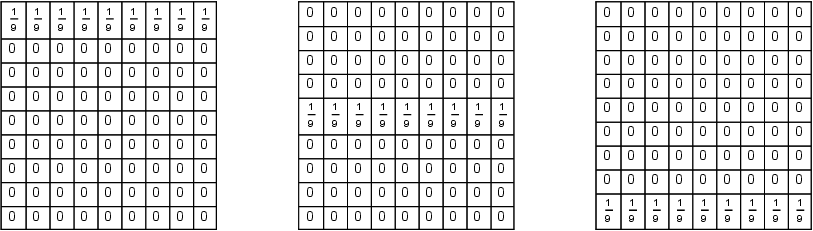}
\caption{DWFs for $\left\vert S_{1}^{(1)}\right\rangle $(Left), $\left\vert
S_{1}^{(2)}\right\rangle $(Middle); $\left\vert S_{1}^{(3)}\right\rangle $%
(Right). They have same 81 values but different distributions.}
\end{figure*}

\begin{figure*}[tbp]
\label{FigWF2} \centering\includegraphics[width=1.9\columnwidth]{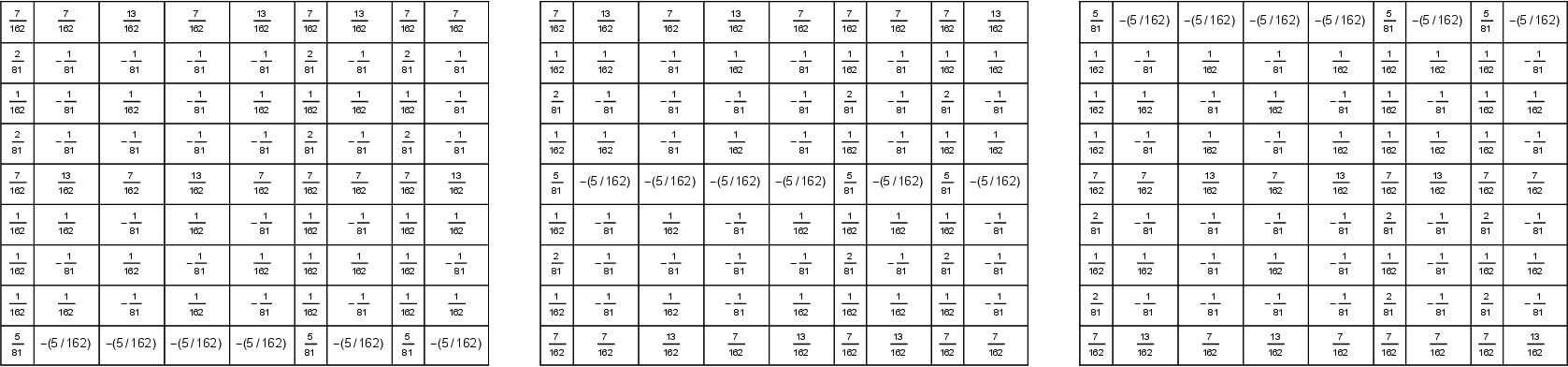}
\caption{DWFs for $\left\vert S_{2}^{(1)}\right\rangle $ (Left); $\left\vert
S_{2}^{(2)}\right\rangle $ (Middle); $\left\vert S_{2}^{(3)}\right\rangle $
(Right). They have same 81 values but different distributions.}
\end{figure*}

\begin{figure*}[tbp]
\label{FigWF3} \centering\includegraphics[width=1.85\columnwidth]{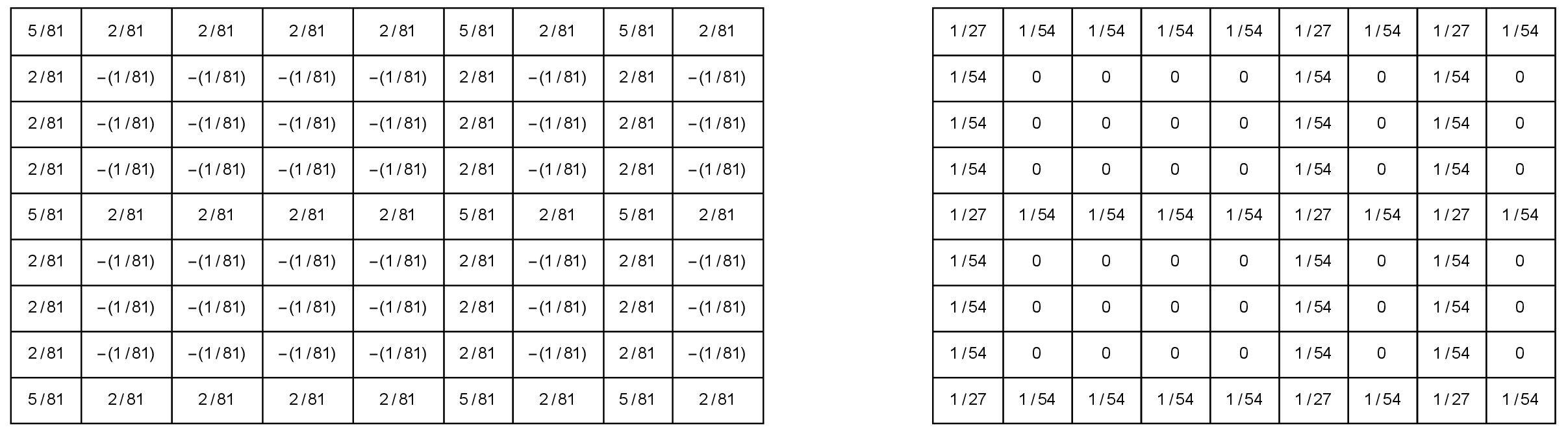}
\caption{DWFs for $\left\vert \Phi _{3}^{+}\right\rangle =\left\vert
S_{3}^{(1)}\right\rangle $ (Left) and $\protect\rho _{noise}=1_{9}/9$
(Right).}
\end{figure*}

As examples, we only plot the DWFs for $\left\vert S_{1}^{(1)}\right\rangle $%
, $\left\vert S_{1}^{(2)}\right\rangle $, $\left\vert
S_{1}^{(3)}\right\rangle $, $\left\vert S_{2}^{(1)}\right\rangle $, $%
\left\vert S_{2}^{(2)}\right\rangle $, $\left\vert S_{2}^{(3)}\right\rangle $%
, $\left\vert S_{3}^{(1)}\right\rangle $ and $\rho _{noise}$ in this work.
The DWFs, for $\left\vert S_{2}^{(4)}\right\rangle $, $\left\vert
S_{2}^{(5)}\right\rangle $, $\left\vert S_{2}^{(6)}\right\rangle $, $%
\left\vert S_{3}^{(2)}\right\rangle $ and other $\rho _{aiso}$s, are not
plotted here.

\subsection{Wigner negativity}

For any quantum state, the Wigner function is normalized. As expected, we
can definitely verify%
\begin{equation}
\sum_{x_{1},z_{1};x_{2},z_{2}\in
\mathbb{Z}
_{3}^{4}}W_{\left( x_{1},z_{1};x_{2},z_{2}\right) }\left( \rho
_{aiso}\right) =1  \label{3.7}
\end{equation}
for any $\rho _{aiso}$. The amount of the Wigner negativity is just minus
the sum of all negative values among the DWF. Hence, the Wigner negativity
of $\rho _{aiso}$ can be calculated by%
\begin{equation}
\mathcal{N}\left( \rho _{aiso}\right) =\frac{1}{2}%
[\sum_{x_{1},z_{1};x_{2},z_{2}\in
\mathbb{Z}
_{3}^{4}}\left\vert W_{\left( x_{1},z_{1};x_{2},z_{2}\right) }\left( \rho
_{aiso}\right) \right\vert -1].  \label{3.8}
\end{equation}%
After making numerical calculation, we easily obtain $\mathcal{N}(\left\vert
S_{1}^{(1)}\right\rangle )=\mathcal{N}(\left\vert S_{1}^{(2)}\right\rangle )=%
\mathcal{N}(\left\vert S_{1}^{(3)}\right\rangle )=0$, $\mathcal{N}%
(\left\vert S_{2}^{(4)}\right\rangle )=\mathcal{N}(\left\vert
S_{2}^{(5)}\right\rangle )=\mathcal{N}(\left\vert S_{2}^{(6)}\right\rangle
)\simeq 0.416975$, $\mathcal{N}(\left\vert S_{2}^{(1)}\right\rangle )=%
\mathcal{N}(\left\vert S_{2}^{(2)}\right\rangle )=\mathcal{N}(\left\vert
S_{2}^{(3)}\right\rangle )=\frac{13}{27}\simeq 0.481481$, $\mathcal{N}%
(\left\vert S_{3}^{(2)}\right\rangle )\simeq 0.421011$, $\mathcal{N}%
(\left\vert S_{3}^{(1)}\right\rangle )=\frac{4}{9}\simeq 0.444444$, and $%
\mathcal{N}\left( \rho _{noise}\right) =0$. Some of these values can be
validated from our plotted DWFs.

Figure 6 depicts the variation of Wigner negativity $\mathcal{N}\left( \rho
_{aiso}\right) $ versus $p$\ for eleven $(\theta ,\phi )$ cases. There are
five curves in this figure. Each curve is illustrated as follows:

(wL1) The first curve corresponds to the cases of $\left( \theta ,\phi
\right) =(\pi /2,0)$, $(\pi /2,\pi /2)$, $\left( 0,\phi \right) $. It
satisfy $\mathcal{N}\left( \rho _{aiso}\right) \equiv 0$ for any $p\in
\lbrack 0,1]$.

(wL2) The second curve corresponds to the cases of $\left( \theta ,\phi
\right) =(\pi /2,\pi /6)$, $(\pi /6,0)$, $(\pi /6,\pi /2)$. It is a
piecewise function line, satisfying $\mathcal{E}\left( \rho _{aiso}\right)
=0 $ in the interval of $0\leq p\lesssim 0.315949$, $\mathcal{N}\left( \rho
_{aiso}\right) \simeq 0.234449p-0.0740741$ in the interval of $%
0.315949\lesssim p\leq 0.535898$, and $\mathcal{N}\left( \rho _{aiso}\right)
=0.787346p-0.37037$ in the interval of $0.535898\lesssim p\leq 1$.

(wL3) The third curve corresponds to the cases of $\left( \theta ,\phi
\right) =(\pi /2,\pi /4)$, $(\pi /4,0)$, $(\pi /4,\pi /2)$. In this curve,
there are three typical points $\left( p,\mathcal{N}\right) =(0.285714,0)$, $%
(0.5,1/18)$, $(1,23/27)$. This curve is a piecewise function line,
satisfying $\mathcal{N}\left( \rho _{aiso}\right) =0$ in the interval of $%
0\leq p\lesssim 0.285714$, $\mathcal{N}\left( \rho _{aiso}\right) \simeq
0.259259p-0.0740741$ in the interval of $0.285714\lesssim p\leq 0.5$, and $%
\mathcal{N}\left( \rho _{aiso}\right) =$ $23p/27-10/27$ in the interval of $%
0.5\lesssim p\leq 1$.

(wL4) The fourth curve corresponds to the case of $\left( \theta ,\phi
\right) =(\arccos (1/\sqrt{3}),\pi /6)$. It is also a piecewise function
line, satisfying $\mathcal{N}\left( \rho _{aiso}\right) =0$ in the interval
of $0\leq p\lesssim 0.463361$, $\mathcal{N}\left( \rho _{aiso}\right) \simeq
0.319725p-0.148148$ in the interval of $0.463361\lesssim p\lesssim 0.500194$%
, $\mathcal{N}\left( \rho _{aiso}\right) \simeq 0.615907p-0.296296$ in the
interval of $0.500194\lesssim p\lesssim 0.61731$, and $\mathcal{N}\left(
\rho _{aiso}\right) \simeq $ $0.0787166p^{2}+0.753744p-0.411381$ in the
interval of $0.61731\lesssim p\leq 1$.

(wL5) The fifth curve corresponds to the case of $\left( \theta ,\phi
\right) =(\arccos (1/\sqrt{3}),\pi /4)$. It is also a piecewise function
line, satisfying $\mathcal{N}\left( \rho _{aiso}\right) =0$ in the interval
of $0\leq p\leq 0.5$ and $\mathcal{N}\left( \rho _{aiso}\right) =8p/9-4/9$
in the interval of $0.5\leq p\leq 1$.

\begin{figure}[tbp]
\label{FigWN1} \centering\includegraphics[width=0.9\columnwidth]{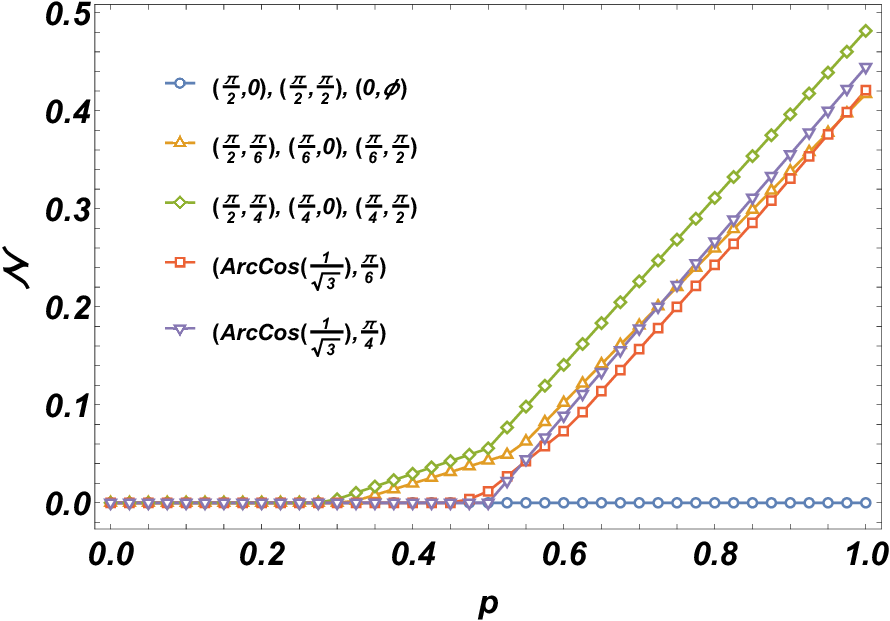}
\caption{$\mathcal{N}\left( \protect\rho _{aiso}\right) $ versus $p$\ for
eleven ($\protect\theta $, $\protect\phi $) cases. There are only five
variation curves. For $p=1$, $\mathcal{N}\left( \protect\rho _{aiso}\right) $
values are$\ 0$, $0.416975$, $0.421011$, $4/9$, and $13/27\ $in order from
small to large.}
\end{figure}

\begin{figure}[tbp]
\label{FigWN2} \centering\includegraphics[width=0.8\columnwidth]{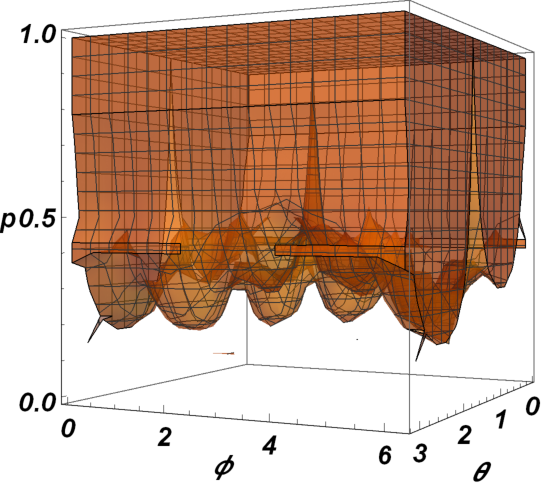}
\caption{Three-dimensional feasibility region with $\mathcal{N}\left(
\protect\rho _{aiso}\right) >0$ showing Wigner negativity in $(\protect%
\theta ,\protect\phi ,p)$\ space. The blank region is that satisfying $%
\mathcal{N}\left( \protect\rho _{aiso}\right) =0$.}
\end{figure}

For different $\left( \theta ,\phi \right) $ cases, the evolution curves of $%
\mathcal{N}\left( \rho _{aiso}\right) $ over $p$ are different. In most
cases, $\mathcal{N}\left( \rho _{aiso}\right) $ value is not equal to $p%
\mathcal{N}\left( \left\vert \psi _{\left( \theta ,\phi \right)
}\right\rangle \left\langle \psi _{\left( \theta ,\phi \right) }\right\vert
\right) +(1-p)\mathcal{N}\left( \rho _{noise}\right) $, except that $%
\mathcal{N}\left( \rho _{aiso}\right) \equiv 0$ when $\left\vert \psi
_{\left( \theta ,\phi \right) }\right\rangle =\left\vert 00\right\rangle $, $%
\left\vert 11\right\rangle $, $\left\vert 22\right\rangle $. Due to the
effects of the noise, the Wigner negativity only occurs when the $p$-value
exceeds a certain value. This can be seen from Fig.7, which plot the
feasibility regions in $\left( \theta ,\phi ,p\right) $\ space, satisfying $%
\mathcal{N}\left( \rho _{aiso}\right) >0$. That is to say, only when $p$%
-value exceeds a certain threshold, it is possible to observe $\mathcal{N}%
\left( \rho _{aiso}\right) >0$ for a given $\left( \theta ,\phi \right) $
case.

In addition, we find that $\mathcal{N}^{\max }\left( \rho _{aiso}\right) =%
\frac{13}{27}$ is found for $\rho _{aiso}\rightarrow \left\vert \psi
_{\left( \theta ,\phi \right) }\right\rangle =\left\vert
S_{2}^{(i)}\right\rangle $ ($i=1,2,3$). It is worth to note that states with
stronger entanglement do not necessarily have greater Wigner negativity. For
instance, although $\mathcal{E}(\left\vert S_{3}^{(1)}\right\rangle )$ is
greater than $\mathcal{E}(\left\vert S_{2}^{(i)}\right\rangle )$ ($i=1,2,3$%
), $\mathcal{N}(\left\vert S_{3}^{(1)}\right\rangle )$ is less than $%
\mathcal{N}(\left\vert S_{2}^{(i)}\right\rangle )$ ($i=1,2,3$).

\section{Bell nonlocality of AITTSs}

As Meyer et al. pointed out\cite{101}, Wigner negativity is necessary for
nonlocality in qudit systems. In this section, we study Bell nonlocality for
the AITTSs by virtue of the CGLMP inequalities. We consider two separated
observers, Alice and Bob. Alice can conduct two different measurements ($%
A_{1}$ and $A_{2}$) with respective three outcomes, i.e., $A_{1}=j$ ($%
j=0,1,2 $) and $A_{2}=k$ ($k=0,1,2$). Similarly, Bob can conduct two
different measurements ($B_{1}$ and $B_{2}$) with respective three outcomes,
i.e., $B_{1}=l$ ($l=0,1,2$) and $B_{2}=m$ ($m=0,1,2$). In order to avoid the
eigenstates degenerate, we further assume that their eigenstates satisfy
\begin{eqnarray}
\left\vert j\right\rangle _{A_{1}} &=&\frac{1}{\sqrt{3}}\sum_{n=0}^{2}\omega
^{n\left( j+\alpha _{1}\right) }\left\vert n\right\rangle _{A},  \label{4.1}
\\
\left\vert k\right\rangle _{A_{2}} &=&\frac{1}{\sqrt{3}}\sum_{n=0}^{2}\omega
^{n(k+\alpha _{2})}\left\vert n\right\rangle _{A},  \label{4.2} \\
\left\vert l\right\rangle _{B_{1}} &=&\frac{1}{\sqrt{3}}\sum_{n=0}^{2}\omega
^{n(-l+\beta _{1})}\left\vert n\right\rangle _{B},  \label{4.3} \\
\left\vert m\right\rangle _{B_{2}} &=&\frac{1}{\sqrt{3}}\sum_{n=0}^{2}\omega
^{n(-m+\beta _{2})}\left\vert n\right\rangle _{B},  \label{4.4}
\end{eqnarray}%
which lead to $\Pi _{A_{1}}^{\left( j\right) }=\left\vert j\right\rangle
_{A_{1}}\left\langle j\right\vert $, $\Pi _{A_{2}}^{\left( k\right)
}=\left\vert k\right\rangle _{A_{2}}\left\langle k\right\vert $, $\Pi
_{B_{1}}^{\left( l\right) }=\left\vert l\right\rangle _{B_{1}}\left\langle
l\right\vert $, and $\Pi _{B_{2}}^{\left( m\right) }=\left\vert
m\right\rangle _{B_{2}}\left\langle m\right\vert $ ($j,k,l,m\in
\mathbb{Z}
_{3}$), respectively.

Now, we let Alice and Bob share the AITTSs. $\allowbreak $The joint
probabilities can be calculated as%
\begin{eqnarray}
P\left( A_{1}=j,B_{1}=l\right) &=&\mathrm{Tr}[(\Pi _{A_{1}}^{\left( j\right)
}\otimes \Pi _{B_{1}}^{\left( l\right) })\rho _{aiso}],  \label{4.5} \\
P\left( A_{1}=j,B_{2}=m\right) &=&\mathrm{Tr}[(\Pi _{A_{1}}^{\left( j\right)
}\otimes \Pi _{B_{2}}^{\left( m\right) })\rho _{aiso}],  \label{4.6} \\
P\left( A_{2}=k,B_{1}=l\right) &=&\mathrm{Tr}[(\Pi _{A_{2}}^{\left( k\right)
}\otimes \Pi _{B_{1}}^{\left( l\right) })\rho _{aiso}],  \label{4.7} \\
P\left( A_{2}=k,B_{2}=m\right) &=&\mathrm{Tr}[(\Pi _{A_{2}}^{\left( k\right)
}\otimes \Pi _{B_{2}}^{\left( m\right) })\rho _{aiso}].  \label{4.8}
\end{eqnarray}%
Further, we use the CGLMP inequality%
\begin{eqnarray}
\mathcal{I}_{3} &\equiv &[P\left( A_{1}=B_{1}\right) +P\left(
B_{1}=A_{2}+1\right)  \notag \\
&&+P\left( A_{2}=B_{2}\right) +P\left( B_{2}=A_{1}\right) ]  \notag \\
&&-[P\left( A_{1}=B_{1}-1\right) +P\left( B_{1}=A_{2}\right)  \notag \\
&&+P\left( A_{2}=B_{2}-1\right) +P\left( B_{2}=A_{1}-1\right) ]  \notag \\
&\leq &2.  \label{4.9}
\end{eqnarray}%
to study the Bell nonlocality for $\rho _{aiso}$ by observing $\mathcal{I}%
_{3}\left( \rho _{aiso}\right) >2$ and setting $(\alpha _{1},\alpha
_{2},\beta _{1},\beta _{2})=(0,1/2,1/4,-1/4)$.

As our references, we give $\mathcal{I}_{3}(\left\vert
S_{2}^{(1)}\right\rangle )=\mathcal{I}_{3}(\left\vert
S_{2}^{(2)}\right\rangle )=\mathcal{I}_{3}(\left\vert
S_{2}^{(3)}\right\rangle )=0$, $\mathcal{I}_{3}(\left\vert
S_{2}^{(4)}\right\rangle )=\mathcal{I}_{3}(\left\vert
S_{2}^{(6)}\right\rangle )\simeq 1$, $\mathcal{I}_{3}(\left\vert
S_{2}^{(1)}\right\rangle )=\mathcal{I}_{3}(\left\vert
S_{2}^{(3)}\right\rangle )\simeq 1.1547$, $\mathcal{I}_{3}(\left\vert
S_{2}^{(5)}\right\rangle )\simeq 1.73205$, $\mathcal{I}_{3}(\left\vert
S_{2}^{(2)}\right\rangle )\simeq 2$, $\mathcal{I}_{3}(\left\vert
S_{3}^{(1)}\right\rangle )\simeq 2.84399$, $\mathcal{I}_{3}(\left\vert
S_{3}^{(1)}\right\rangle )\simeq 2.87293$, and $\mathcal{I}_{3}\left( \rho
_{noise}\right) =0$. Surprisedly, for those Sn-2 states, we find $\mathcal{I}%
_{3}(\left\vert S_{2}^{(1)}\right\rangle )=\mathcal{I}_{3}(\left\vert
S_{2}^{(3)}\right\rangle )\neq \mathcal{I}_{3}(\left\vert
S_{2}^{(2)}\right\rangle )$ and $\mathcal{I}_{3}(\left\vert
S_{2}^{(4)}\right\rangle )=\mathcal{I}_{3}(\left\vert
S_{2}^{(6)}\right\rangle )\neq \mathcal{I}_{3}(\left\vert
S_{2}^{(5)}\right\rangle )$. These results are different from those in
studying $\mathcal{E}$ and $\mathcal{N}$.

\begin{figure}[tbp]
\label{FigBN1} \centering\includegraphics[width=0.9\columnwidth]{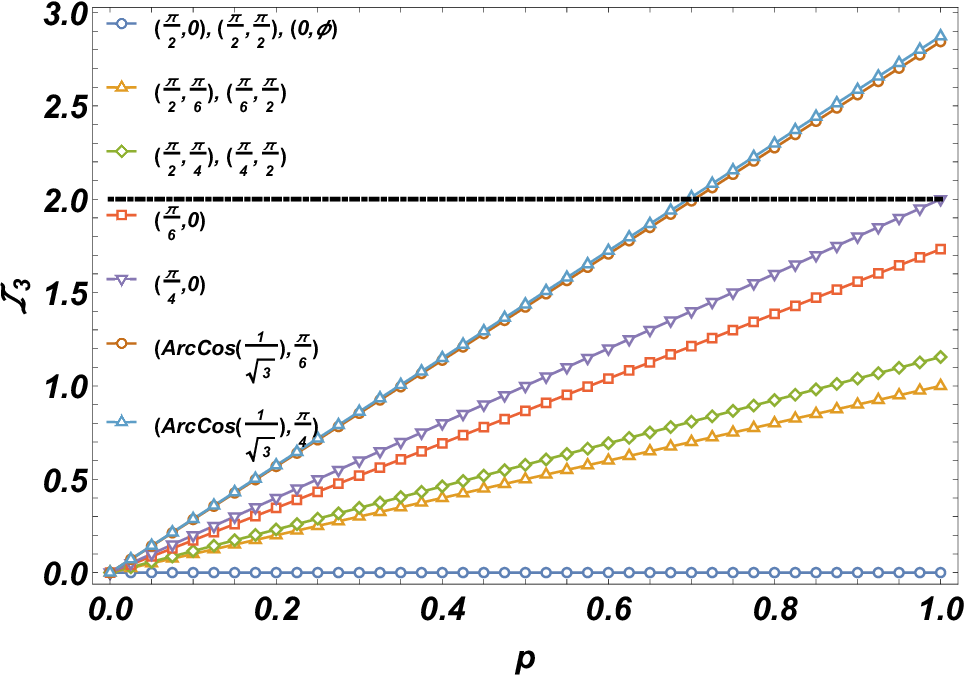}
\caption{$\mathcal{I}_{3}\left( \protect\rho _{aiso}\right) $ versus $p$\
for eleven ($\protect\theta $, $\protect\phi $) cases. For $p=1$, $\mathcal{I%
}_{3}$ values are$\ 0$, $1$, $1.1547$, $1.73205$, $2$, $2.84399$, $2.87293$
in sequence. Each line is a straight line.}
\end{figure}

\begin{figure}[tbp]
\label{FigBN2} \centering\includegraphics[width=0.8\columnwidth]{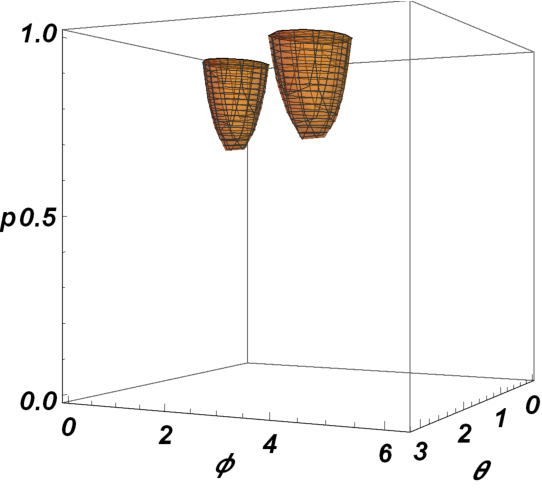}
\caption{Three-dimensional feasibility region with $\mathcal{I}_{3}\left(
\protect\rho _{aiso}\right) >2$ showing Bell nonlocality in $(\protect\theta %
,\protect\phi ,p)$\ space. The blank region is that satisfying $\mathcal{I}%
_{3}\left( \protect\rho _{aiso}\right) \leq 2$.}
\end{figure}

Fig.8 depicts $\mathcal{I}_{3}\left( \rho _{aiso}\right) $s as functions of $%
p$\ for eleven $(\theta ,\phi )$cases. There are seven lines in this figure.
Each line is illustrated as follows:

(bL1) The first line corresponds to the cases of $\left( \theta ,\phi
\right) =(\pi /2,0)$, $(\pi /2,\pi /2)$, $\left( 0,\phi \right) $. It
satisfy $\mathcal{I}_{3}\left( \rho _{aiso}\right) \equiv 0$ for any $p\in
\lbrack 0,1]$.

(bL2) The second line corresponds to the cases of $\left( \theta ,\phi
\right) =(\pi /2,\pi /6)$, $(\pi /6,\pi /2)$. It is a straight line
piecewise satisfy $\mathcal{I}_{3}\left( \rho _{aiso}\right) =p$ in the
whole $p$ range.

(bL3) The third curve corresponds to the cases of $\left( \theta ,\phi
\right) =(\pi /2,\pi /4)$, $(\pi /4,\pi /2)$. It is a straight line
piecewise satisfy $\mathcal{I}_{3}\left( \rho _{aiso}\right) =1.1547p$ in
the whole $p$ range.

(bL4) The fourth line corresponds to the cases of $\left( \theta ,\phi
\right) =(\pi /6,0)$. It is a straight line piecewise satisfy $\mathcal{I}%
_{3}\left( \rho _{aiso}\right) =1.73205p$ in the whole $p$ range.

(bL5) The fifth curve corresponds to the cases of $\left( \theta ,\phi
\right) =(\pi /4,0)$. It is a straight line piecewise satisfy $\mathcal{I}%
_{3}\left( \rho _{aiso}\right) =2p$ in the whole $p$range.

(bL6) The sixth curve corresponds to the case of $\left( \theta ,\phi
\right) =(\arccos (1/\sqrt{3}),\pi /6)$. It is a straight line piecewise
satisfy $\mathcal{I}_{3}\left( \rho _{aiso}\right) =2.84399p$ in the whole $%
p $ range.

(bL7) The seventh curve corresponds to the case of $\left( \theta ,\phi
\right) =(\arccos (1/\sqrt{3}),\pi /4)$. It is a straight line piecewise
satisfy $\mathcal{I}_{3}\left( \rho _{aiso}\right) =2.87293p$ in the whole $%
p $ range.

It is worth noting that we have $\mathcal{I}_{3}\left( \rho _{aiso}\right) =p%
\mathcal{I}_{3}\left( \left\vert \psi _{\left( \theta ,\phi \right)
}\right\rangle \right) $ for arbitrarily determined $\left( \theta ,\phi
\right) $ case. That is, $\mathcal{I}_{3}\left( \rho _{aiso}\right) $ is a
linear function of $p$ ($\in \lbrack 0,1]$) with slope value $\mathcal{I}%
_{3}\left( \left\vert \psi _{\left( \theta ,\phi \right) }\right\rangle
\right) $. Interestingly, we can also draw the conclusion $\mathcal{I}%
_{3}\left( \rho _{aiso}\right) =p\mathcal{I}_{3}\left( \left\vert \psi
_{\left( \theta ,\phi \right) }\right\rangle \right) +(1-p)\mathcal{I}%
_{3}\left( \rho _{noise}\right) $.

Theoretically, Bell nonlocality can be witnessed by observing $\mathcal{I}%
_{3}\left( \rho _{aiso}\right) >2$, i.e. the violation of Eq.(\ref{4.9}).
From above numerical results, we immediately know that there is the
possibility of $\mathcal{I}_{3}\left( \rho _{aiso}\right) >2$ if and only if
$\left\vert \psi _{\left( \theta ,\phi \right) }\right\rangle $\ is the Sn=3
state. In Fig.9, we plot the feasibility regions in $\left( \theta ,\phi
,p\right) $\ space showing Bell nonlocality. Only in the range of $%
0.686141\lesssim p\leq 1$, we can choose proper $\left( \theta ,\phi \right)
$ values to satisfy $\mathcal{I}_{3}\left( \rho _{aiso}\right) >2$.
Moreover, the optional $\left( \theta ,\phi \right) $ region area will
decrease\ as $p$ decreases until $p\approx 0.686141$. In other words, no
matter how you choose $\left( \theta ,\phi \right) $ if $0\leq p\lesssim
0.686141$, it is impossible to ensure that this inequality $\mathcal{I}%
_{3}\left( \rho _{aiso}\right) >2$ holds true. Here, I would like to remind
everyone that $\mathcal{I}_{3}\left( \rho _{aiso}\right) $ values may be
less than zero for some parameter $\left( \theta ,\phi \right) $ regions,
but $\mathcal{I}_{3}\left( \rho _{aiso}\right) $ values are absolutely
impossible to be less than $-2$ for any parameter $\left( \theta ,\phi
\right) $ regions.

In addition, we find that the maximum value of $\mathcal{I}_{3}\left( \rho
_{aiso}\right) $ is $\mathcal{I}_{3}^{\max }=2.91485$, which is positioned
at $\left( \theta ,\phi ,p\right) \simeq \left( 0.906006,0.67002,1\right) \ $%
or $\left( 2.23559,3.81161,1\right) $. This is an astonishing result for us
because of $\mathcal{I}_{3}^{\max }\left( \rho _{aiso}\right) \geq \mathcal{I%
}_{3}^{\max }\left( \rho _{iso}\right) =\mathcal{I}_{3}(\left\vert \Phi
_{3}^{+}\right\rangle )\simeq 2.87293$. That is to say, the maximally
entangled state is not the maximally Bell-nonlocality state. In fact, there
are a lot of $\rho _{aiso}$s, whose $\mathcal{I}_{3}$ values are greater
than $\mathcal{I}_{3}(\left\vert \Phi _{3}^{+}\right\rangle )$. As shown in
Fig.10, two feasibility regions satisfying $\mathcal{I}_{3}\left( \rho
_{aiso}\right) >\mathcal{I}_{3}(\left\vert \Phi _{3}^{+}\right\rangle )$ are
distributed in the interval of $0.985618\lesssim p\leq 1$, together with
proper $\left( \theta ,\phi \right) $ values satisfying $0.8112\lesssim
\theta \lesssim 1.002$, $0.5388\lesssim \phi \lesssim 0.7996$, or $%
2.141\lesssim \theta \lesssim 2.330$, $3.681\lesssim \phi \lesssim 3.941$.

\begin{figure}[tbp]
\label{Fig10ab} \centering\includegraphics[width=0.8\columnwidth]{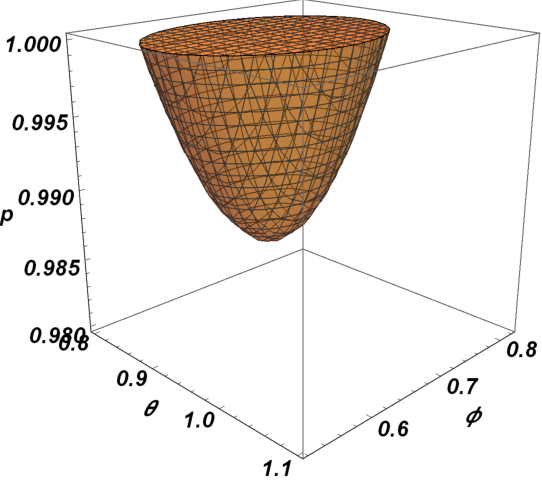}
\centering\includegraphics[width=0.8\columnwidth]{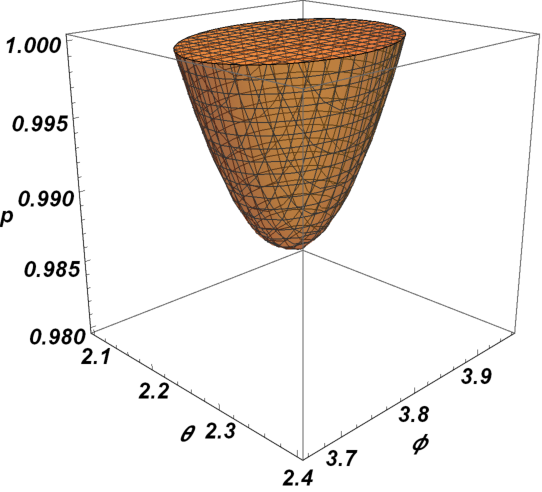}
\caption{ Two\ feasibility regions satisfying $\mathcal{I}_{3}\left( \protect%
\rho _{aiso}\right) >2.87293$ in $(\protect\theta ,\protect\phi ,p)$\ space.}
\end{figure}

\section{Conclusions and discussions}

In summary, we introduced the AITTSs, and then explored their properties,
including entanglement (see index $\mathcal{E}$), Wigner negativity (see
index $\mathcal{N}$) and Bell nonlocality (see index $\mathcal{I}_{3}$). For
all these properties, we tried our best to obtain their respective
analytical and numerical results.

As one extremal case of $\rho _{aiso}$ with $p=0$, we always had $\mathcal{E}%
\left( \rho _{noise}\right) \equiv 0$, $\mathcal{N}\left( \rho
_{noise}\right) \equiv 0$, and $\mathcal{I}_{3}\left( \rho _{noise}\right)
\equiv 0$. As another extremal case of $\rho _{aiso}$ with $p=1$, we know
that $\mathcal{E}\left( \left\vert \psi _{\left( \theta ,\phi \right)
}\right\rangle \right) $, $\mathcal{N}\left( \left\vert \psi _{\left( \theta
,\phi \right) }\right\rangle \right) $, and $\mathcal{I}_{3}\left(
\left\vert \psi _{\left( \theta ,\phi \right) }\right\rangle \right) $ are
determined by $\left\vert \psi _{\left( \theta ,\phi \right) }\right\rangle $
in itself. So, in the end of this paper, we specialize in analyzing the
influence of parameters $\left( \theta ,\phi \right) $ on these three
properties of $\left\vert \psi _{\left( \theta ,\phi \right) }\right\rangle $%
. Figure 11 depicts the contours for $\mathcal{E}\left( \left\vert \psi
_{\left( \theta ,\phi \right) }\right\rangle \right) $, $\mathcal{N}\left(
\left\vert \psi _{\left( \theta ,\phi \right) }\right\rangle \right) $, and $%
\mathcal{I}_{3}\left( \left\vert \psi _{\left( \theta ,\phi \right)
}\right\rangle \right) $\ in $\left( \theta ,\phi \right) $ plains. From
Fig.11(a), we see that $\mathcal{E}\left( \left\vert \psi _{\left( \theta
,\phi \right) }\right\rangle \right) $\ is a periodic function of $\theta $\
with period $\pi $\ and a periodic function of $\phi $\ with period $\pi /4$%
. Compared with $\mathcal{E}\left( \left\vert \psi _{\left( \theta ,\phi
\right) }\right\rangle \right) $, the periodic features of $\mathcal{N}%
\left( \left\vert \psi _{\left( \theta ,\phi \right) }\right\rangle \right) $
are gone, as shown in Fig.11(b). By observing $\mathcal{I}_{3}\left(
\left\vert \psi _{\left( \theta ,\phi \right) }\right\rangle \right) $ in
Fig.11(c), we find that $\mathcal{I}_{3}\left( \left\vert \psi _{\left(
\theta ,\phi \right) }\right\rangle \right) $\ values are in the range of $%
[-2,2.91485]$. Specially, when $\left( \theta ,\phi ,p\right) =\left( \pi
/4,\pi ,1\right) $, the minimal $\mathcal{I}_{3}^{\min }=-2$ is found,
corresponding to $\left\vert \psi _{\left( \theta ,\phi \right)
}\right\rangle =\frac{1}{\sqrt{2}}(\left\vert 22\right\rangle -\left\vert
00\right\rangle )$.

\begin{figure*}[tbp]
\label{Fig11} \centering\includegraphics[width=1.8\columnwidth]{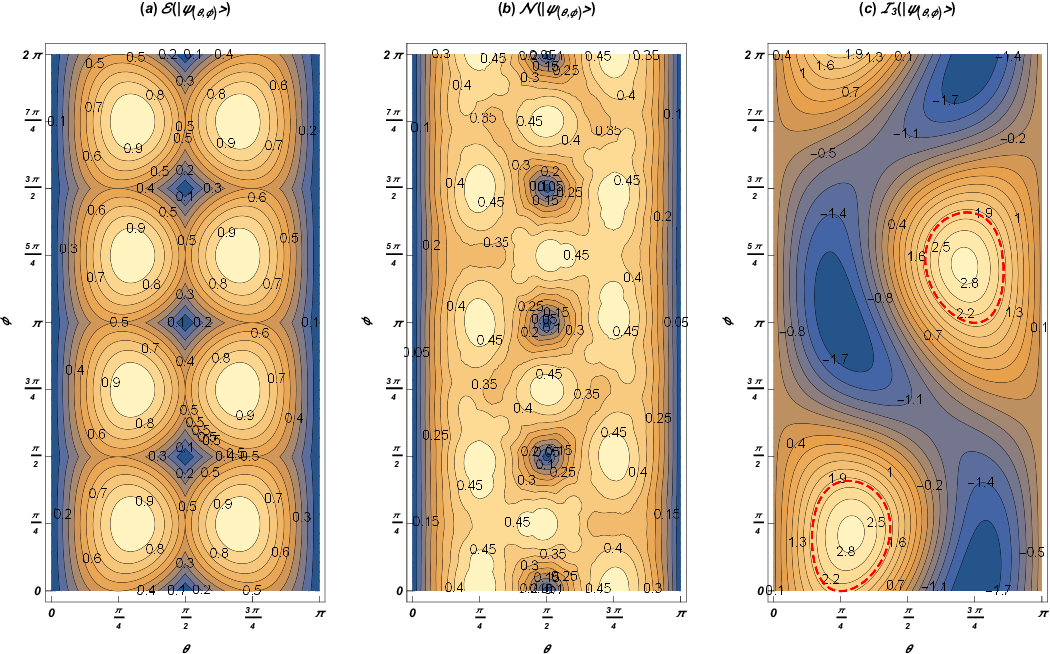}
\caption{Contourplots of (a) $\mathcal{E}\left( \left\vert \protect\psi %
_{\left( \protect\theta ,\protect\phi \right) }\right\rangle \right) $, (b) $%
\mathcal{N}\left( \left\vert \protect\psi _{\left( \protect\theta ,\protect%
\phi \right) }\right\rangle \right) $, and (c) $\mathcal{I}_{3}\left(
\left\vert \protect\psi _{\left( \protect\theta ,\protect\phi \right)
}\right\rangle \right) $\ in $\left( \protect\theta ,\protect\phi \right) $
plain\ space. It is necessary to remind in\ sub-figure (c) that (i) the red
dashed lines correspond to $\mathcal{I}_{3}\left( \left\vert \protect\psi %
_{\left( \protect\theta ,\protect\phi \right) }\right\rangle \right) =2$;
(ii) For some $\left( \protect\theta ,\protect\phi \right) $ regions, $%
\mathcal{I}_{3}$ values are less than 0.}
\end{figure*}

If $\left\vert \psi _{\left( \theta ,\phi \right) }\right\rangle $ was the
Sn=1 state, then we always have $\mathcal{E}\left( \rho _{aiso}\right)
\equiv 0$, $\mathcal{N}\left( \rho _{aiso}\right) \equiv 0$, and $\mathcal{I}%
_{3}\left( \rho _{aiso}\right) \equiv 0$. While $\left\vert \psi _{\left(
\theta ,\phi \right) }\right\rangle $ was not the Sn=1 states, then we found
that (i) $\mathcal{E}\left( \rho _{aiso}\right) $ was not a linear function
of $p$; (ii) $\mathcal{N}\left( \rho _{aiso}\right) $ was also not a linear
function of $p$; (iii) but, $\mathcal{I}_{3}\left( \rho _{aiso}\right) $ was
a linear function of $p$. Although $\rho _{aiso}$ was the mixture between $%
\left\vert \psi _{\left( \theta ,\phi \right) }\right\rangle $\ and $\rho
_{noise}$ (see Eq.(\ref{1.2})), we concluded that (i) The equality of $%
\mathcal{E}\left( \rho _{aiso}\right) =p\mathcal{E}\left( \left\vert \psi
_{\left( \theta ,\phi \right) }\right\rangle \right) +(1-p)\mathcal{E}\left(
\rho _{noise}\right) $ may not necessarily hold true; (ii) The equality of $%
\mathcal{N}\left( \rho _{aiso}\right) =p\mathcal{N}\left( \left\vert \psi
_{\left( \theta ,\phi \right) }\right\rangle \right) +(1-p)\mathcal{N}\left(
\rho _{noise}\right) $ may not necessarily hold true; (iii) But the equality
of $\mathcal{I}_{3}\left( \rho _{aiso}\right) =p\mathcal{I}_{3}\left(
\left\vert \psi _{\left( \theta ,\phi \right) }\right\rangle \right) +(1-p)%
\mathcal{I}_{3}\left( \rho _{noise}\right) $ may always hold true.

Finally, we summarize several key results as follows:

(1) There is no decisive relationship between these three properties. If a
quantum state has the highest entanglement, its negativity may not
necessarily be the highest, and its Bell non-locality may not necessarily be
the strongest. This point can be verified from $\mathcal{E}$, $\mathcal{N}$,
and $\mathcal{I}_{3}$ of state $\left\vert \Phi _{3}^{+}\right\rangle $. The
maximally entangled state is not the maximally Wigner-negativity state and
the maximally Bell-nonlocality state.

(2) A quantum pure state with a larger Schmidt number does not necessarily
have a greater Wigner negativity. This point can be verified from $\mathcal{N%
}(\left\vert S_{2}^{(1)}\right\rangle )>\mathcal{N}(\left\vert
S_{3}^{(1)}\right\rangle )$.

(3) It is the effects of the noise that three properties will exhibit only
when $p$ exceeds a certain threshold. This point can be seen from our
numerical results. Of course, the optimal properties of $\rho _{aiso}$ are
those of $\left\vert \psi _{\left( \theta ,\phi \right) }\right\rangle $
corresponding to $p=1$, without the effects of the noise.

\begin{acknowledgments}
This paper was supported by the National Natural Science Foundation of China
(Grant No. 12465004).
\end{acknowledgments}

\section*{Appendix}

\textbf{Appendix A}:\ Pauli matrices related with qutrit system

The bases are set as $\left\vert 0\right\rangle =\left(
\begin{array}{ccc}
1 & 0 & 0%
\end{array}%
\right) ^{T}$, $\left\vert 1\right\rangle =\left(
\begin{array}{ccc}
0 & 1 & 0%
\end{array}%
\right) ^{T}$, $\left\vert 2\right\rangle =\left(
\begin{array}{ccc}
0 & 0 & 1%
\end{array}%
\right) ^{T}$, $\left\langle 0\right\vert =\left(
\begin{array}{ccc}
1 & 0 & 0%
\end{array}%
\right) $, $\left\langle 1\right\vert =\left(
\begin{array}{ccc}
0 & 1 & 0%
\end{array}%
\right) $, and $\left\langle 2\right\vert =\left(
\begin{array}{ccc}
0 & 0 & 1%
\end{array}%
\right) $, for qutrit system with dimensionality $d=3$ and $\omega =e^{i%
\frac{2\pi }{3}}$. Related Pauli matrices are created via repeated matrix
multiplication $X^{x}Z^{z}$ ($x$, $z\in
\mathbb{Z}
_{3}$)\cite{56} as
\begin{subequations}
\begin{align}
X^{0}Z^{0}& =\left(
\begin{array}{ccc}
1 & 0 & 0 \\
0 & 1 & 0 \\
0 & 0 & 1%
\end{array}%
\right) ,X^{0}Z^{1}=\left(
\begin{array}{ccc}
1 & 0 & 0 \\
0 & \omega & 0 \\
0 & 0 & \omega ^{2}%
\end{array}%
\right) ,  \tag{A.1} \\
X^{0}Z^{2}& =\left(
\begin{array}{ccc}
1 & 0 & 0 \\
0 & \omega ^{2} & 0 \\
0 & 0 & \omega%
\end{array}%
\right) ,X^{1}Z^{0}=\left(
\begin{array}{ccc}
0 & 0 & 1 \\
1 & 0 & 0 \\
0 & 1 & 0%
\end{array}%
\right) ,  \tag{A.2} \\
X^{1}Z^{1}& =\left(
\begin{array}{ccc}
0 & 0 & \omega ^{2} \\
1 & 0 & 0 \\
0 & \omega & 0%
\end{array}%
\right) ,X^{1}Z^{2}=\left(
\begin{array}{ccc}
0 & 0 & \omega \\
1 & 0 & 0 \\
0 & \omega ^{2} & 0%
\end{array}%
\right) ,  \tag{A.3} \\
X^{2}Z^{0}& =\left(
\begin{array}{ccc}
0 & 1 & 0 \\
0 & 0 & 1 \\
1 & 0 & 0%
\end{array}%
\right) ,X^{2}Z^{1}=\left(
\begin{array}{ccc}
0 & \omega & 0 \\
0 & 0 & \omega ^{2} \\
1 & 0 & 0%
\end{array}%
\right) ,  \tag{A.4} \\
X^{2}Z^{2}& =\left(
\begin{array}{ccc}
0 & \omega ^{2} & 0 \\
0 & 0 & \omega \\
1 & 0 & 0%
\end{array}%
\right) .  \tag{A.5}
\end{align}%
According to Eq.(\ref{3.3}) and using above matrices, we can obtain the
matrix in Eq.(\ref{3.4}).

\textbf{Appendix B}:\ Matrices for $\Pi _{A_{1}}^{\left( j\right) }$, $\Pi
_{A_{2}}^{\left( k\right) }$, $\Pi _{B_{1}}^{\left( l\right) }$, and $\Pi
_{B_{2}}^{\left( m\right) }$

(1) Matrix of $\Pi _{A_{1}}^{\left( j\right) }$ is
\end{subequations}
\begin{subequations}
\begin{equation}
\Pi _{A_{1}}^{\left( j\right) }=\frac{1}{3}\allowbreak \left(
\begin{array}{ccc}
1 & \omega ^{-j-\alpha _{1}} & \omega ^{-2j-2\alpha _{1}} \\
\omega ^{j+\alpha _{1}} & 1 & \omega ^{-j-\alpha _{1}} \\
\omega ^{2j+2\alpha _{1}} & \omega ^{j+\alpha _{1}} & 1%
\end{array}%
\right) ,  \tag{B.1}
\end{equation}

(2) Matrix of $\Pi _{A_{2}}^{\left( k\right) }$ is

\end{subequations}
\begin{subequations}
\begin{equation}
\Pi _{A_{2}}^{\left( k\right) }=\frac{1}{3}\allowbreak \left(
\begin{array}{ccc}
1 & \omega ^{-k-\alpha _{2}} & \omega ^{-2k-2\alpha _{2}} \\
\omega ^{k+\alpha _{2}} & 1 & \omega ^{-k-\alpha _{2}} \\
\omega ^{2k+2\alpha _{2}} & \omega ^{k+\alpha _{2}} & 1%
\end{array}%
\right) ,  \tag{B.2}
\end{equation}

(3) Matrix of $\Pi _{B_{1}}^{\left( l\right) }$ is

\end{subequations}
\begin{subequations}
\begin{equation}
\Pi _{B_{1}}^{\left( l\right) }=\frac{1}{3}\left(
\begin{array}{ccc}
1 & \omega ^{l-\beta _{1}} & \omega ^{2l-2\beta _{1}} \\
\omega ^{\beta _{1}-l} & 1 & \omega ^{l-\beta _{1}} \\
\omega ^{2\beta _{1}-2l} & \omega ^{\beta _{1}-l} & 1%
\end{array}%
\right) ,  \tag{B.3}
\end{equation}

(4) Matrix of $\Pi _{B_{2}}^{\left( m\right) }$ is

\end{subequations}
\begin{subequations}
\begin{equation}
\Pi _{B_{2}}^{\left( m\right) }=\frac{1}{3}\left(
\begin{array}{ccc}
1 & \omega ^{m-\beta _{2}} & \omega ^{2m-2\beta _{2}} \\
\omega ^{\beta _{2}-m} & 1 & \omega ^{m-\beta _{2}} \\
\omega ^{2\beta _{2}-2m} & \omega ^{\beta _{2}-m} & 1%
\end{array}%
\right) .  \tag{B.4}
\end{equation}%
\textbf{Appendix C}:\ Details of each term in Eq.(\ref{4.9})

(1) For $P\left( A_{1}=B_{1}\right) $, we have

\end{subequations}
\begin{subequations}
\begin{align}
P\left( A_{1}=B_{1}\right) & =P\left( A_{1}=0,B_{1}=0\right)  \notag \\
& +P\left( A_{1}=1,B_{1}=1\right)  \notag \\
& +P\left( A_{1}=2,B_{1}=2\right) ,  \tag{C.1}
\end{align}

(2) For $P\left( B_{1}=A_{2}+1\right) $, we have
\end{subequations}
\begin{subequations}
\begin{align}
P\left( B_{1}=A_{2}+1\right) & =P\left( A_{2}=0,B_{1}=1\right)  \notag \\
& +P\left( A_{2}=1,B_{1}=2\right)  \notag \\
& +P\left( A_{2}=2,B_{1}=0\right) ,  \tag{C.2}
\end{align}

(3) For $P\left( A_{2}=B_{2}\right) $, we have
\end{subequations}
\begin{subequations}
\begin{align}
P\left( A_{2}=B_{2}\right) & =P\left( A_{2}=0,B_{2}=0\right)  \notag \\
& +P\left( A_{2}=1,B_{2}=1\right)  \notag \\
& +P\left( A_{2}=2,B_{2}=2\right) ,  \tag{C.3}
\end{align}

(4) For $P\left( B_{2}=A_{1}\right) $, we have
\end{subequations}
\begin{subequations}
\begin{align}
P\left( B_{2}=A_{1}\right) & =P\left( A_{1}=0,B_{2}=0\right)  \notag \\
& +P\left( A_{1}=1,B_{2}=1\right)  \notag \\
& +P\left( A_{1}=2,B_{2}=2\right) ,  \tag{C.4}
\end{align}

(5) For $P\left( A_{1}=B_{1}-1\right) $, we have
\end{subequations}
\begin{subequations}
\begin{align}
P\left( A_{1}=B_{1}-1\right) & =P\left( A_{1}=2,B_{1}=0\right)  \notag \\
& +P\left( A_{1}=0,B_{1}=1\right)  \notag \\
& +P\left( A_{1}=1,B_{1}=2\right) ,  \tag{C.5}
\end{align}

(6) For $P\left( B_{1}=A_{2}\right) $, we have
\end{subequations}
\begin{subequations}
\begin{align}
P\left( B_{1}=A_{2}\right) & =P\left( A_{2}=0,B_{1}=0\right)  \notag \\
& +P\left( A_{2}=1,B_{1}=1\right)  \notag \\
& +P\left( A_{2}=2,B_{1}=2\right) ,  \tag{C.6}
\end{align}

(7) For $P\left( A_{2}=B_{2}-1\right) $, we have
\end{subequations}
\begin{subequations}
\begin{align}
P\left( A_{2}=B_{2}-1\right) & =P\left( A_{2}=2,B_{2}=0\right)  \notag \\
& +P\left( A_{2}=0,B_{2}=1\right)  \notag \\
& +P\left( A_{2}=1,B_{2}=2\right) ,  \tag{C.7}
\end{align}

(8) For $P\left( B_{2}=A_{1}-1\right) $, we have
\end{subequations}
\begin{subequations}
\begin{align}
P\left( B_{2}=A_{1}-1\right) & =P\left( A_{1}=0,B_{2}=2\right)  \notag \\
& +P\left( A_{1}=1,B_{2}=0\right)  \notag \\
& +P\left( A_{1}=2,B_{2}=1\right) .  \tag{C.8}
\end{align}

\end{subequations}


\bigskip

\begin{thebibliography}{999}
\bibitem{1} E. Chitambar and G. Gour, Quantum resource theories, Rev. Mod.
Phys. 91, 025001 (2019).

\bibitem{2} U. Chabaud and M. Walschaers, Resources for Bosonic Quantum
Computational Advantage, Phys. Rev. Lett. 130, 090602 (2023).

\bibitem{3} A. Mari, K. Kieling, B. Melholt Nielsen, E. S. Polzik, and J.
Eisert, Directly Estimating Nonclassicality, Phys. Rev. Lett. 106, 010403
(2011).

\bibitem{4} C. Gehrke, J. Sperling, and W. Vogel, Quantification of
nonclassicality, Phys. Rev. A 86, 052118 (2012).

\bibitem{5} N. Killoran, F.\thinspace E.\thinspace S. Steinhoff, and
M.\thinspace B. Plenio, Converting Nonclassicality into Entanglement, Phys.
Rev. Lett. 116, 080402 (2016).

\bibitem{6} S. Ryl, J. Sperling, E. Agudelo, M. Mraz, S. Kohnke, B. Hage,
and W. Vogel, Unified nonclassicality criteria, Phys. Rev. A 92, 011801(R)
(2015).

\bibitem{7} B. Kuhn and W. Vogel, Quantum non-Gaussianity and quantification
of nonclassicality, Phys. Rev. A 97, 053823 (2018).

\bibitem{8} Q. Zhuang, P. W. Shor, and J. H. Shapiro, Resource theory of
non-Gaussian operations, Phys. Rev. A 97, 052317 (2018).

\bibitem{9} F. Albarelli, M. G. Genoni, M. G. A. Paris, and A. Ferraro,
Resource theory of quantum non-Gaussianity and Wigner negativity, Phys. Rev.
A 98, 052350 (2018).

\bibitem{10} R. Takagi and Q. Zhuang, Convex resource theory of
non-Gaussianity, Phys. Rev. A 97, 062337 (2018).

\bibitem{11} Z. S. Zhang, et al., Entanglement-based quantum information
technology: a tutorial, Adv. Opt. Photon. 16, 60-162 (2024).

\bibitem{12} N. Friis, G. Vitagliano, M. Malik, and M. Huber, Entanglement
certification from theory to experiment, Nat. Rev. Phys. 1, 72-87 (2019).

\bibitem{13} M. Erhard, M. Krenn, and A. Zeilinger, Advances in
high-dimensional quantum entanglement, Nat. Rev. Phys. 2, 365-381 (2020).

\bibitem{14} Y. Fan, C. Jia, and L. Qiu, Quantum steering as resource of
quantum teleportation, Phys. Rev. A 106, 012433 (2022).

\bibitem{15} R. Uola, A.C.S. Costa, H. C. Nguyen, and O. Guhne, Quantum
steering, Rev. Mod. Phys. 92, 015001 (2020).

\bibitem{16} R. Gallego and L. Aolita, Resource Theory of Steering, Phys.
Rev. X 5, 041008 (2015).

\bibitem{17} C. Vieira, R. Ramanathan and A. Cabello, Test of the physical
significance of Bell non-locality, Nature Communications 16, 4390 (2025).

\bibitem{18} A. Tavakoli, A. Pozas-Kerstjens, M. X. Luo, M. O. Renou, Bell
nonlocality in networks, Rep. Prog. Phys. 85, 056001 (2022).

\bibitem{19} U. Chabaud, P. E. Emeriau, and F. Grosshans, Witnessing Wigner
Negativity, Quantum 5, 471 (2021).

\bibitem{20} V. Veitch, C. Ferrie, D. Gross and J. Emerson, Negative
quasi-probability as a resource for quantum computation, New J. Phys. 14,
113011 (2012).

\bibitem{21} C. Budroni, A. Cabello, O. Guhne, M. Kleinmann, and J. Larsson,
Kochen-Specker contextuality, Rev. Mod. Phys. 94, 045007 (2022).

\bibitem{22} F. Shahandeh Contextuality of General Probabilistic Theories,
PRX Quantum 2, 010330 (2021).

\bibitem{23} A. Grudka, K. Horodecki, M. Horodecki, P. Horodecki, R.
Horodecki, P. Joshi, W. Klobus, and A. Wojcik, Quantifying Contextuality,
Phys. Rev. Lett. 112, 120401 (2014).

\bibitem{24} R. I. Booth, U. Chabaud, and P. E. Emeriau, Contextuality and
Wigner Negativity Are Equivalent for Continuous-Variable Quantum
Measurements, Phys. Rev. Lett. 129, 230401 (2022)

\bibitem{25} N. Delfosse, P. A. Guerin, J. Bian, and R. Raussendorf, Wigner
Function Negativity and Contextuality in Quantum Computation on Rebits,
Phys. Rev. X 5, 021003 (2015).

\bibitem{26} R. Raussendorf, D. E. Browne, N. Delfosse, C. Okay, and J.
Bermejo-Vega, Contextuality and Wigner-function negativity in qubit quantum
computation, Phys. Rev. A 95, 052334 (2017).

\bibitem{27} M. Walschaers, C. Fabre, V. Parigi, and N. Treps, Entanglement
and Wigner Function Negativity of Multimode Non-Gaussian States, Phys. Rev.
Lett. 119, 183601 (2017).

\bibitem{28} A. Cabello, Converting Contextuality into Nonlocality, Phys.
Rev. Lett. 127, 070401 (2021).

\bibitem{29} K. Svozil, Converting nonlocality into contextuality, Phys.
Rev. A 110, 012215 (2024).

\bibitem{30} M. Plavala and O. Guhne, Contextuality as a Precondition for
Quantum Entanglement, Phys. Rev. Lett. 132, 100201 (2024).

\bibitem{31} P. A. M. Dirac, The Principles of Quantum Mechanics, Oxford
University Press, 1958.

\bibitem{32} G. Auletta, M. Fortunato, and G. Parisi, Quantum Mechanics,
Cambridge University Press, 2009.

\bibitem{33} E. Wigner, On the Quantum Correction for Thermodynamic
Equilibrium, Phys. Rev. 40, 749-759 (1932).

\bibitem{34} T. Tilma, M. J. Everitt, J. H. Samson, W. J. Munro, and K.
Nemoto, Wigner Functions for Arbitrary Quantum Systems, Phys. Rev. Lett.
117, 180401 (2016).

\bibitem{35} W. P. Schleich, Quantum Optics in Phase Space, Berlin:
Wiley-VCH, 2001.

\bibitem{36} U. Chabaud, P. E. Emeriau, and F. Grosshans, Witnessing Wigner
Negativity, Quantum 5, 471 (2021).

\bibitem{37} Y. Xiang, S. Liu, J. Guo, Q. Gong, N. Treps, Q. He, and M.
Walschaers, Quantification of Wigner Negativity Remotely Generated via
Einstein-Podolsky-Rosen Steering, npj Quantum Information 8, 21 (2022).

\bibitem{38} M. Walschaers, On Quantum Steering and Wigner Negativity,
Quantum 7, 1038 (2023).

\bibitem{39} A. Kenfack and K. Zyczkowski, Negativity of the Wigner function
as an indicator of non-classicality, J. Opt. B: Quantum Semiclassical Opt.
6, 396 (2004).

\bibitem{40} M. Walschaers, Non-gaussian quantum states and where to find
them, PRX Quantum 2, 030204 (2021).

\bibitem{41} F. X. Sun, Y. Q. Fang, Q. Y. He, Y. Q. Liu, Generating optical
cat states via quantum interference of multi-path free-electron-photons
interactions, Sci. Bull. 68, 1366-1371 (2023).

\bibitem{42} D. J. Wineland, Superposition, entanglement, and raising
Schrodinger's cat, Ann. Phys. 525, 739-752 (2013).

\bibitem{43} A. Mari and J. Eisert, Positive Wigner Functions Render
Classical Simulation of Quantum Computation Efficient, Phys. Rev. Lett. 109,
230503 (2012).

\bibitem{44} U. Chabaud and M. Walschaers, Resources for Bosonic Quantum
Computational Advantage, Phys. Rev. Lett. 130 090602 (2023).

\bibitem{45} R. L. Hudson, When is the Wigner quasi-probability density
non-negative? Rep. Math. Phys. 6, 249-252 (1974).

\bibitem{46} F. Soto and P. Claverie, When is the Wigner function of
multidimensional systems nonnegative? J. Math. Phys. 24, 97 (1983).

\bibitem{47} D. Gross, Hudson's theorem for finite-dimensional quantum
systems, J. Math. Phys. 47, 122107 (2006).

\bibitem{48} N. Dangniam, Y. G. Han, and H. Zhu, Optimal verification of
stabilizer states, Phys. Rev. Research 2, 043323 (2020).

\bibitem{49} N. Koukoulekidis and D. Jennings, Constraints on magic state
protocols from the statistical mechanics of Wigner negativity, npj Quantum
Information 8, 42 (2022).

\bibitem{50} A. Casaccino, E. F. Galvao, and S. Severini, Extrema of
discrete Wigner functions and applications, Phys. Rev. A 78, 022310 (2008).

\bibitem{51} K. S. Gibbons, M. J. Hoffman, and W. K. Wootters, Discrete
phase space based on finite fields, Phys. Rev. A 70, 062101 (2004).

\bibitem{52} E. F. Galvao, Discrete Wigner functions and quantum
computational speedup, Phys. Rev. A 71, 042302 (2005).

\bibitem{53} L. Kocia and P. Love, Discrete Wigner formalism for qubits and
noncontextuality of Clifford gates on qubit stabilizer states, Phys. Rev. A
96, 062134 (2017).

\bibitem{54} W. K. Wootters, Interpreting symplectic linear transformations
in a two-qubit phase space, Int. J. Quan. Inf. 22(05), 2440014 (2024).

\bibitem{55} L. K. Antonopoulos, D. G. Lewis, J. Davis, N. Funai, and N. C.
Menicucci, Grand Unification of All Discrete Wigner Functions on $d\times d$
Phase Space, arXiv: 2503.09353 (2025).

\bibitem{56} P. Kok and B, W. Lovett, Introduction to Optical Quantum
Information Processing, New York: Cambridge Unoversity Press, 2010.

\bibitem{57} M. Erhard, M. Krenn, and A. Zeilinger, Advances in
high-dimensional quantum entanglement, Nat. Rev. Phys. 2, 365-381 (2020).

\bibitem{58} Y. Li, Y. Xiang, X. D. Yu, H. Chau Nguyen, O. Guhne, Q. Y. He,
Randomness Certification from Multipartite Quantum Steering for Arbitrary
Dimensional Systems, Phys. Rev. Lett. 132, 080201 (2024).

\bibitem{59} H. J. Kimble, The quantum internet, Nature 453 (7198),
1023-1030 (2008).

\bibitem{60} S. Esposito, The quantum internet - the second quantum
revolution, Contemp. Phys. 63 (4), 328-328 (2022).

\bibitem{61} S. Wehner, D. Elkouss, and R. Hanson, Quantum internet: A
vision for the road ahead, Science 362, eaam9288 (2018).

\bibitem{62} Y. Xiang, F. X. Sun, Q. Y. He, Q. H. Gong, Advances in
multipartite and high-dimensional Einstein-Podolsky-Rosen steering,
Fundamental Research 1, 99-101 (2021)

\bibitem{63} D. Cozzolino, B. Da Lio, D. Bacco, and L. K. Oxenlowe,
High-Dimensional Quantum Communication: Benefits, Progress, and Future
Challenges, Adv. Quan. Tech. 2, 1900038 (2019

\bibitem{64} N. D'Alessandro, C. R. Carceller, and A. Tavakoli, Semidefinite
Relaxations for High-Dimensional Entanglement in the Steering Scenario,
Phys. Rev. Lett. 134, 090802 (2025).

\bibitem{65} N. K. H. Li, M. Huber and N. Friis, High-dimensional
entanglement witnessed by correlations in arbitrary bases, npj Quantum
Information 11, 50 (2025).

\bibitem{66} S. Morelli, M. Huber, and A. Tavakoli, Resource-Efficient
High-Dimensional Entanglement Detection via Symmetric Projections, Phys.
Rev. Lett. 131, 170201 (2023).

\bibitem{67} Z. Huang, L. Maccone, A. Karim, C. Macchiavello, R. J. Chapman,
and A. Peruzzo, High-dimensional entanglement certification, Sci. Rep. 6,
27637 (2016).

\bibitem{68} J. Wang et al., Multidimensional quantum entanglement with
large-scale integrated optics, Science 10.1126/science.arr7053 (2018).

\bibitem{69} O. Lib, S. Liu, R. Shekel, Q. He, M. Huber, Y. Bromberg, G.
Vitagliano, Experimental certification of high-dimensional entanglement with
randomized measurements, arxiv: 2412.04643 (2024).

\bibitem{70} S. H. Liu, M. Fadel, Q. Y. He, M. Huber, G. Vitagliano,
Bounding entanglement dimensionality from the covariance matrix, Quantum 8,
1236 (2024).

\bibitem{71} J. Bell, On the Einstein podolsky rosen paradox, Physics, 1,
195-199, (1964).

\bibitem{72} M. Karczewski, G. Scala, A. Mandarino, A. B. Sainz and M.
Zukowski, Avenues to generalising Bell inequalities, J. Phys. A: Math.
Theor. 55 384011 (2022).

\bibitem{73} J. Kaniewski, I. Supic, J. Tura, F. Baccari, A. Salavrakos, R.
Augusiak, Maximal nonlocality from maximal entanglement and mutually
unbiased bases, and self-testing of two-qutrit quantum systems, Quantum 3,
198 (2019)

\bibitem{74} O. Veltheim and E. Keski-Vakkuri, Optimizing Quantum
Measurements by Partitioning Multisets of Observables, Phys. Rev. Lett. 134,
030801 (2025).

\bibitem{75} M. Pandit, A. Barasi\'{n}ski, I. Marton, T. Vertesi and W.
Laskowski, Optimal tests of genuine multipartite nonlocality, New J. Phys.
24 123017 (2022).

\bibitem{76} A. Barasinski, Exploring nonlocal correlations in arbitrarily
high-dimensional systems, New J. Phys. 26 113013 (2024).

\bibitem{77} A. Fonseca, A. de Rosier, T. Vertesi, W. Laskowski, and F.
Parisio, Survey on the Bell nonlocality of a pair of entangled qudits, Phys.
Rev. A 98, 042105 (2018).

\bibitem{78} D. Collins, N. Gisin, N. Linden, S. Massar, and S. Popescu,
Bell Inequalities for Arbitrarily High-Dimensional Systems, Phys. Rev. Lett.
88, 040404 (2002).

\bibitem{79} I. Supic and J. Bowles, Self-testing of quantum systems: a
review, Quantum 4, 337 (2020)

\bibitem{80} J. Kaniewski, I. Supic, J. Tura, F. Baccari, A. Salavrakos, and
R. Augusiak, Maximal nonlocality from maximal entanglement and mutually
unbiased bases, and self-testing of two-qutrit quantum systems, Quantum 3,
198 (2019).

\bibitem{81} Z. Cai, C. Ren, T. Feng, X. Zhou, and J. Chen, A review of
quantum correlation sharing: The recycling of quantum correlations triggered
by quantum measurements, Phys. Rep. 1098 1 (2025).

\bibitem{82} C. Zhang, Y. Li, X. M. Hu, Y. Xiang, C. F. Li, G. C. Guo, J.
Tura, Q. Gong, Q. He, and B. H. Liu, Randomness versus Nonlocality in
Multiple-Input and Multiple-Output Quantum Scenario, Phys. Rev. Lett. 134
090201 (2025).

\bibitem{83} C. Brukner, M. Zukowski, and A. Zeilinger, Quantum
Communication Complexity Protocol with Two Entangled Qutrits, Phys. Rev.
Lett. 89, 197901 (2002).

\bibitem{84} M.\thinspace A. Yurtalan, J. Shi, M. Kononenko, A. Lupascu, and
S. Ashhab, Implementation of a Walsh-Hadamard Gate in a Superconducting
Qutrit, Phys. Rev. Lett. 125, 180504 (2020).

\bibitem{85} F. Turro, I. A. Chernyshev, R. Bhaskar, and M. Illa, Qutrit and
qubit circuits for three-flavor collective neutrino oscillations, Phys. Rev.
D 111, 043038 (2025).

\bibitem{86} D. Amaro-Alcala, B. C. Sanders, and H. de Guise, Benchmarking
of universal qutrit gates, Phys. Rev. A 109, 012621 (2024).

\bibitem{87} A. B. Klimov, R. Guzman, J. C. Retamal, and C. Saavedra, Qutrit
quantum computer with trapped ions, Phys. Rev. A 67, 062313 (2003).

\bibitem{88} J. Gruca, W. Laskowski, and M. Zukowski, Nonclassicality of
pure two-qutrit entangled states, Phys. Rev. A 85, 022118 (2012).

\bibitem{89} S. Roy, A. Kumari, S. Mal, and A. Sen(De), Robustness of
higher-dimensional nonlocality against dual noise and sequential
measurements, Phys. Rev. A 109, 062227 (2024).

\bibitem{90} R. Lifshitz, Noise-Robust Self-Testing: Detecting Non-Locality
in Noisy Non-Local Inputs, Master's Thesis, Weizmann Institute of Science,
arXiv: 2505.13537 (2025).

\bibitem{91} J. M. Liang, S. Imai, S. Liu, S. M. Fei, O. Guhne, Q. He, Real
randomized measurements for analyzing properties of quantum states, arXiv:
2411.06013 (2024).

\bibitem{92} D. Kaszlikowski, P. Gnacinski, M. Zukowski, W. Miklaszewski,
and A. Zeilinger, Violations of Local Realism by Two Entangled N-Dimensional
Systems Are Stronger than for Two Qubits, Phys. Rev. Lett. 85, 4418 (2000).

\bibitem{93} T. Durt, D. Kaszlikowski, and M. Zukowski, Violations of local
realism with quantum systems described by N-dimensional Hilbert spaces up to
N=16, Phys. Rev. A 64, 024101 (2001).

\bibitem{94} B. M. Terhal and P. Horodecki, Schmidt number for density
matrices, Phys. Rev. A 61, 040301(R) (2000).

\bibitem{95} A. Sanpera, D. Bru$\beta $ and M. Lewenstein, Schmidt-number
witnesses and bound entanglement, Phys. Rev. A 63, 050301(R) (2001).

\bibitem{96} J. Sperling and W. Vogel, The Schmidt number as a universal
entanglement measure, Phys. Scr. 83 045002 (2011)

\bibitem{97} H.-F. Wang and S. M. Fei, Schmidt number criterion via
symmetric measurements, arXiv:2505.02297 (2005).

\bibitem{98} G. Vidal and R. F. Werner, Computable measure of entanglement,
Phys. Rev. A 65, 032314 (2002).

\bibitem{99} M. Horodecki, P. Horodecki, and R. Horodecki, Separability of
mixed states: Necessary and sufficient conditions, Phys. Lett. A 223, 1-8
(1996).

\bibitem{100} N. Delfosse, C. Okay, J. Bermejo-Vega, D. E. Browne and R.
Raussendorf, Equivalence between contextuality and negativity of the Wigner
function for qudits, New J. Phys. 19, 123024 (2017).

\bibitem{101} U. I. Meyer, I. Supic, D. Markham, and F. Grosshans, Qudit
Clauser-Horne-Shimony-Holt Inequality and Nonlocality from Wigner
Negativity, arXiv: 2405.14367 (2024).
\end{thebibliography}
\end{document}